\documentclass[aps,prc,twocolumn,superscriptaddress]{revtex4-1}

\usepackage{graphicx}
\usepackage{amsmath}
\usepackage{lineno}

\usepackage[colorlinks,
linkcolor=blue,
anchorcolor=blue,
urlcolor=red,
citecolor=blue]{hyperref}

\begin{document}
\title{Revisiting the variational two-particle reduced density matrix for nuclear systems}


\author{J. G. Li}
\affiliation{School of Physics,  and   State Key  Laboratory  of  Nuclear  Physics   and  Technology, Peking University, Beijing  100871, China}
\author{N. Michel}
\affiliation{Institute of Modern Physics, Chinese Academy of Sciences, Lanzhou 730000, China}
\affiliation{School of Nuclear Science and Technology, University of Chinese Academy of Sciences, Beijing 100049, China}
\author{W. Zuo}
\affiliation{Institute of Modern Physics, Chinese Academy of Sciences, Lanzhou 730000, China}
\affiliation{School of Nuclear Science and Technology, University of Chinese Academy of Sciences, Beijing 100049, China}
\author{F. R. Xu}\email[]{frxu@pku.edu.cn}
\affiliation{School of Physics,  and   State Key  Laboratory  of  Nuclear  Physics   and  Technology, Peking University, Beijing  100871, China}


\date{\today}

\begin{abstract}

In most nuclear many-body methods, observables are calculated using many-body wave functions explicitly. The variational two-particle reduced density matrix method is one of the few exceptions to the rule. Ground-state energies of both  closed-shell and open-shell nuclear systems can indeed be evaluated by minimizing a constrained linear functional of the two-particle reduced density matrix.  However, it has virtually never been used in nuclear theory, because nuclear ground states were found to be well overbound, contrary to those of atoms and molecules. Consequently, we introduced new constraints in the nuclear variational two-particle reduced density matrix method, developed recently for atomic and molecular systems. Our calculations then show that this approach can provide a proper description of nuclear systems where only valence neutrons are included. For the nuclear systems where both neutrons and protons are active, however, the energies obtained with the variational two-particle reduced density matrix method are still overbound.  The possible reasons for the noticed discrepancies and solutions to this problem will be discussed.

\end{abstract}


\maketitle
\section{Introduction}

In all areas of modern theoretical quantum mechanics, which comprises nuclear physics, quantum chemistry, and condensed matter physics, many-body physics consist in the study of correlated systems. They are induced by the two-body interactions present between the fundamental particles building the many-body system. The main problem arising when one aims at solving the many-body problem in quantum mechanics is practical: the number of the occupied orbitals at basis level increases linearly with system size, which causes the number of configurations necessary to precisely represent the correlated many-body wave function to increase \textit{exponentially}.

In nuclear physics, most of the many-body methods are based on the use of an explicit many-body wave function, which is a solution of the many-body Schr{\"o}dinger equation. Most popular methods are based on shell-model \cite{RevModPhys.77.427,RevModPhys.92.015002}, variational quantum Monte Carlo \cite{RevModPhys.87.1067} and valence space in-medium similarity renormalization group methods \cite{HERGERT2016165,annurev-nucl-101917-021120}.
Other methods have been developed to recapture the features of the many-body wave function with objects of reduced size, such as with the coupled-cluster theory \cite{hagen2016emergent}, where one uses a similarity transformed Hamiltonian acting on a simple reference state, for example a Hartree-Fock state, using an exponentiated operator. However, the coupled-cluster theory is typically limited to the use of singles-and-doubles approximations \cite{hagen2016emergent}.
Many-body wave functions are not strictly necessary in order to calculate observables of many-body quantum systems. Indeed, all one-body and two-body observables can be exactly calculated using the two-particle reduced density matrix (2RDM) \cite{PhysRev.100.1579,PhysRevLett.76.1039,RevModPhys.35.668,PhysRevLett.93.213001,PhysRevA.71.062503,PhysRevA.72.032510,PhysRevA.74.032501,PhysRevA.80.032508,PhysRevA.80.042109,PhysRevLett.108.263002,PhysRevLett.108.213001,PhysRevLett.117.153001}.
In fact, the number of variables of the 2RDM functional scales only \textit{polynomially} with the number of basis orbitals, which is far less than the exponential increase of basis configurations in configuration interaction methods.

The 2RDM method is equivalent to solving the Schr$\ddot{\text{o}}$dinger equation in order to obtain the many-body wave functions.  The ground-state energy is indeed a well-defined linear functional of the Hamiltonian  matrix and 2RDM elements in the variational 2RDM method.
For a $N$-body system, the Hamiltonian can be written as :
\begin{eqnarray}\label{equation1}
   \hat{H} = \sum_{ij} t_{ij}a_{i}^{\dagger}a_{j} + \frac{1}{2}\sum_{ijkl} V_{ij;kl} a_{i}^{\dagger}a_{j}^{\dagger} a_{l} a_{k}.
\end{eqnarray}
And, the energy of the system can be expressed as \cite{PhysRevA.80.032508}:
\begin{eqnarray}\label{equation2}
   E(\Gamma) =  \text{Tr}~ \Gamma H^{(2)} = \frac{1}{4} \sum_{ijkl} \Gamma_{ij;kl} H^{(2)}_{ij;kl},
\end{eqnarray}
where the 2RDM elements appear :
\begin{eqnarray}\label{equation3}
   \Gamma_{ij;kl} = \langle \Psi^N | a_{i}^{\dagger}a_{j}^{\dagger} a_{l} a_{k}| \Psi^N \rangle,
\end{eqnarray}
along with the reduced Hamiltonian two-body matrix elements :
\begin{eqnarray}\label{equation4}
 H_{ij;kl}^{(2)}  &=&\frac{1}{N-1}(\delta_{ik}t_{jl} - \delta_{il} t_{jk} - \delta_{jk}t_{il} + \delta_{jl} t_{ik})  \nonumber \\
  &+& V_{ij;kl},
\end{eqnarray}
where $t$ is the one-body kinetic energy operator and $V$ is the two-body interaction, containing the Coulomb force for atomic systems and the Coulomb+nuclear force for nuclear systems, and $N$ is the number of electrons or nucleons.

2RDM was introduced by L$\ddot{\text{o}}$wdin, Mayer, and Coleman in the early days of many-body quantum physics \cite{PhysRev.97.1474,PhysRev.100.1579,RevModPhys.35.668}.
One hereby exploits the fact that a trial 2RDM can be considered as a variable to obtain the ground-state energy without using many-body wave functions. The 2RDM can then be obtained from a minimization procedure, arising from the variational principle, and hence being equivalent to solving the full many-body Schr{\"o}dinger equation \cite{PhysRev.97.1474}.
However, contrary to initial expectations, the trial 2RDM, obtained by minimizing energy and subject only to wave function normalization, is very different from the exact 2RDM arising from many-body wave function eigenstates. In fact, even in weakly correlated atoms and molecules, correlation energy is too large by orders of magnitude \cite{2RDM_Coleman_intro}.
It then became clear that the 2RDM obtained by energy minimization must be constrained by additional rules (or conditions) in order to reproduce the physical 2RDM. Coleman described these necessary and sufficient conditions as $N\text{-}representability$ conditions \cite{RevModPhys.35.668}, i.e.,~that the 2RDM functional is associated to a physical many-body wave function. However, it was demonstrated that, in practice, demanding for an 2RDM functional to fulfill exactly $N\text{-}representability$ conditions is as difficult as solving the many-body Schr{\"o}dinger equation \cite{Garrod_Percus}. Approximate methods thus had to be devised for 2RDM-based methods to be useful in practice.

Research on 2RDM methods has been systematically done for 50 years on atomic and molecular electronic systems \cite{RevModPhys.35.668,PhysRevA.71.062503,Zhao2004}, which allowed to identify the most important nontrivial conditions needed to constrain the 2RDM functional.
These conditions  consist in positive semidefinite operators, whose expectation value thus has to be non-negative when evaluated with a 2RDM. Moreover, they can be systematically arranged in a hierarchy where each level yields an increasingly tighter lower bound on the exact ground-state energy \cite{PhysRevA.80.042109,PhysRevLett.106.083001,PhysRevLett.117.153001,PhysRevLett.108.263002,ANDERSON201322,doi:10.1063/1.1360199}.
However, limitations in optimization software and computer resources prevented the practical development of the 2RDM method for several years.
Significant progress has been made over the last decade in 2RDM methods, as the possible use of powerful computers allowed to devise numerical schemes otherwise impossible to apply in practice \cite{doi:10.1063/1.1360199,PhysRevLett.76.1039,PhysRevLett.93.213001,PhysRevA.80.032508,PhysRevA.80.042109,PhysRevLett.108.263002,PhysRevLett.108.213001,PhysRevLett.117.153001,PhysRevLett.97.143002,PhysRevA.102.030802,PhysRevA.100.032509,acs.jctc.7b00366}.
Two approaches have been developed to calculate the ground state 2RDM : (i) variational 2RDM with energy minimization \cite{PhysRevLett.93.213001,PhysRevLett.108.213001,PhysRevLett.117.153001} and (ii) non-variational approaches based on solutions of the contracted Schr$\ddot{\text{o}}$dinger equation \cite{PhysRevLett.76.1039,PhysRevLett.97.143002,PhysRevA.102.030802}. In the present work, we focus on the variational 2RDM method.  The variational 2RDM minimization problem with positivity constraints can be treated by a special class of optimization techniques known as semidefinite programming  \cite{PhysRevLett.93.213001,PhysRevA.80.032508,PhysRevA.80.042109,PhysRevLett.99.243002,PhysRevLett.106.083001}. Note that these optimization schemes also have applications in control theory \cite{10.1007/978-3-319-12583-1_1}, combinatorial optimization \cite{ant_colony} and finance \cite{ZHANG20061407}.

The first 2RDM calculations of nuclear systems were devised in 1970s \cite{MIHAILOVIC1975221}. However, the obtained ground-state energies were typically overbound by several MeV, and even sometimes tens of MeV.
It was then clear that the variational 2RDM method could not compete with configuration interaction methods effected in small model spaces, for example, with which tremendous success was obtained in the description of $p$ \cite{Cohen_Kurath}, $sd$ \cite{PhysRevLett.20.1112} and $psd$ crossed-shell nuclei \cite{PhysRevLett.21.39,PhysRevLett.23.983}.
Many works \cite{PhysRevLett.102.202501,TDDM1,TDDM2,TDDM3,barton2021nuclear} have been tried to implement the 2RDM technique using a truncated time-dependent Bogoliubov-Born-Green-Kirkwood-Yvon approach together with an adiabatic method for the ground state of nuclear systems. However, there has been no variational 2RDM calculation of nuclear system for a long time. 
Nevertheless, shell-model approaches have not solved the many-body problem in the nuclear chart. Indeed, even with the most powerful machines, shell-model calculations in large model spaces cannot be performed at present without model space truncations \cite{RevModPhys.77.427}. The situation is even more drastic for heavy nuclei, where only mean-field based methods can be applied in systematic calculations \cite{erler2012limits,PhysRevC.82.024313,PhysRevC.68.054312,RevModPhys.75.121,PhysRevC.82.024313,PhysRevC.97.044313}.
Consequently, one can still find interest in the use of the variational 2RDM method even in our era of intense numerical calculations.

The 2RDM method is a powerful many-body method for the description of atomic and molecular systems  \cite{PhysRevLett.76.1039,PhysRevLett.93.213001,PhysRevA.80.032508,PhysRevA.80.042109,PhysRevLett.108.263002,PhysRevLett.108.213001,PhysRevLett.117.153001,PhysRevLett.97.143002,PhysRevA.102.030802}. The precision of 2RDM calculations is, in fact, comparable with coupled-cluster with singles-and-doubles approximations for these weakly correlated systems. Furthermore, it has been shown that variational 2RDM calculations can capture many-electron excitations in the calculations of atomic systems \cite{PhysRevLett.93.213001}. The variational 2RDM has also been successfully applied in the context of the pairing model \cite{Dukelsky_RDM}. In the present paper, we will employ the variational 2RDM for nuclear systems.  The \textit{partial} 3-particle $T_1$ and $T_2^\prime$ conditions,  are first considered in the variational 2RDM calculations for the nuclear systems.  Systematic calculations are performed using variational 2RDM with different conditions for nuclear systems.

This paper is constructed as follows:  we will introduce the variational 2RDM method in Sec.(\ref{Method}) at theoretical level. The used \textit{N-representability} conditions, fundamental to determine the 2RDM by energy minimization, will be described. The standard 2-positivity and  3-positivity \textit{N-representability} conditions developed for the study of molecular systems and which we will use afterwards in nuclei will be presented. The computational method used for constrained energy minimization, based on the semidefinite optimization boundary point method, will be shortly dealt with. Results will be shown in Sec.(\ref{Results}), where ground states of nuclear Hamiltonians based on cluster-orbital shell-model (COSM) and standard shell-model (SM) frameworks are considered. The energies obtained with both 2RDM and exact full configuration interaction methods will be compared. One will see that the quality of results largely differ according to the nuclear system considered. The influences of the 2-positivity and  3-positivity \textit{N-representability} conditions on nuclear ground states will be emphasized for that matter. Plans for future use of the variational 2RDM will then be depicted. Conclusion will be made afterwards.

\section{Method} \label{Method}

In this section, we introduce the theoretical details underlying the variational 2RDM framework. Standard \textit{N-representability} conditions, as well as the energy optimization procedure, belonging to the class of semidefinite optimization boundary point methods, will be presented and extended to the case of nuclear systems.

\subsection{Variational 2RDM method}

The energy of a many-body system is a linear functional of the reduced density matrix. For a two-body Hamiltonian, the ground-state energy $E_g$ can be written as
\begin{eqnarray}\label{Eg}
   E_g = \text{Tr} [H~{^2\Gamma}] = \text{Tr}[H^{(2)}~{^2 \Gamma}],
\end{eqnarray}
where $H^{(2)}$ is the two-body reduced Hamiltonian and $^2 \Gamma$ is two-particle reduced density matrix. The definition of the  $H^{(2)}$ and $^2 \Gamma$ operators can be found in Eqs. (\ref{equation3}) and (\ref{equation4}). The only difference from atomic and molecular systems is that one has both valence proton and neutron occupied states \cite{MIHAILOVIC1975221}. In nuclear physics, we typically treat the many-body problem by including the nuclear interaction as a two-body operator (the three-body force is present by way of a normal-order approach in realistic calculations \cite{PhysRevC.76.034302}).  The 2RDM is then sufficient to calculate all the observables related up to a one-body or two-body operator.

An arbitrary 2RDM is, in general, not associated to a many-body wave function, so that the use of an underconstrained 2RDM typically leads to overbound ground states \cite{2RDM_Coleman_intro,Verstichel_PhD,MIHAILOVIC1975221}. Conversely, except for the trivial case of non-interacting particles,
the minimization of the energy in Eq. (\ref{Eg}) with respect to the 2RDM provides with the exact ground-state energy if the energy minimization is effected by constraining Hamiltonian expectation values over all possible $N$-representable 2RDM:
\begin{eqnarray}\label{Eg_min}
   E_g = \mathop{\text{min}} \limits_{^2\Gamma \in ^2\mathcal{D_N}}\text{Tr}[H^{(2)}~{^2 \Gamma}],
\end{eqnarray}
where $^2\mathcal{D_N}$ contains all the 2RDM functionals generated by a many-body wave function. In order to show that a given 2RDM belongs to $^2\mathcal{D_N}$, one would have to check that the expectation value of all combinations of creation and annihilation operators is positive or equal to zero. Clearly, this is as difficult as solving the many-body Schr{\"o}dinger equation, so that Eq. (\ref{Eg_min}) is only of theoretical interest. Therefore, in practice, one replaces $^2\mathcal{D_N}$ by the set of 2RDM functionals verifying a subset of \textit{N-representability} conditions, which consist in the \textit{N-representability} conditions, developed in molecular 2RDM variational approaches \cite{Garrod_Percus,Erdahl1978,Zhao2004}.

\subsection{\textit{N-representability} conditions and many-body quantum numbers} \label{N_rep}

In 1932, von Neumann  showed that a general matrix is a density matrix only if it fulfills the following conditions \cite{Von_density}: the matrix must be (i) Hermitian, (ii) normalized (fixed trace), (iii) antisymmetric (fermions) or symmetric (bosons) under particle exchange, and (iv) positive semidefinite to yield non-negative probabilities to find particles. Variational minimization of the energy in Eq. (\ref{Eg_min}) with respect to the 2RDM constrained by von Neumann conditions thus provides with the exact ground-state energy. The positive semidefinite character of $\Gamma$ is called the $P$ condition \cite{Garrod_Percus,RevModPhys.35.668}.

The 2RDM can be mapped onto the 1RDM through operator contractions:
\begin{eqnarray}\label{1RDM}
   \rho^{(\nu)}_{ij} = \frac{1}{A-1} \sum_{k}\Gamma_{ik;jk},
\end{eqnarray}
where $\nu$ stands for proton or neutron and $A$ is the number of nucleons.
Consequently, one obtains that the 1RDM is positive semidefinite, i.e.,~$\Gamma \geq 0 \Rightarrow \rho^{(\nu)} \geq  0$. Moreover, the density normalization implies that the trace of the proton (neutron) 1RDM must be equal to the proton (neutron) number:
\begin{eqnarray}\label{particle number}
\text{Tr} \rho^{(\nu)} = N_\nu.
\end{eqnarray}
For simplicity, one will no longer state in the following whether a creation/annihilation operator is related to proton or neutron states, even though this must be effected in practice.

Additional ensembles of \textit{N-representability} conditions can be obtained by considering the positivity of operators related to the 2RDM.
Different representations of $p$RDMs can be easily defined using second quantization \cite{PhysRevLett.108.263002}. For example, let us consider the metric (or overlap) matrices $M$ defined by:
\begin{eqnarray}\label{metric}
   M_{ij} = \langle\Phi_i|\Phi_j\rangle  = \langle \Psi | C_iC_j^{\dagger}|\Psi\rangle,
\end{eqnarray}
from the set of basis functions
\begin{eqnarray}\label{Phi}
   \langle\Phi_i | = \langle \Psi | C_i,
\end{eqnarray}
where each $C_i$ is a product of $p$ creation and/or annihilation operator. When $C_i$  are products of $p$ creation operators, the metric matrix in Eq. (\ref{metric}) is the $p$RDM.  Added to that, one can generate $p$ additional metric matrices if the $C_i$ become products of $p$ second-quantized operators having different numbers of creation and annihilation operators.
These $p+1$ metric matrices must be positive semidefinite \cite{PhysRevLett.108.263002,2RDM_Coleman,2RDM_Mazzi}.  The demanded positivity of all the considered metric matrices of order $p$ is donated as the $p$-positivity condition.

Two independent classes of matrices can be constructed to impose the 1-positivity condition, defined as
\begin{eqnarray}\label{rho}
   \rho_{ij} = \langle\Phi_{i}^{\rho}|\Phi_{j}^{\rho}\rangle  = \langle \psi | a_i^{\dagger}a_j|\psi\rangle, ~~\rho \geq 0,
\end{eqnarray}
\begin{eqnarray}\label{p}
   q_{ij} = \langle\Phi_{i}^{q}|\Phi_{j}^{q}\rangle  = \langle \psi |a_i a_j^{\dagger}|\psi\rangle,~~ q \geq 0,
\end{eqnarray}
where $|\Phi_{j}^{\rho}\rangle$ and $|\Phi_{j}^{q}\rangle$ are $(N-1)$- and $(N+1)$- particle basis function, respectively. $\rho_{ij}$ corresponds to the 1RDM, whereas $q_{ij}$ can be expressed as a linear function of $\rho_{ij}$ by using the anticommutation relations:
\begin{eqnarray}\label{1-positive}
   q_{ij} = \delta_{ij} - \rho_{ij},
\end{eqnarray}
The 1-positivity condition, embodied by the two metric matrices ($\rho_{ij}$  and $q_{ij}$), implies that the eigenvalues of a fermionic 1RDM are between 0 and 1.

Three independent metric matrices can also be generated from the operators $C_i$ out of the 2-positivity condition. They consist of the $P$-,$Q$-, and $G$- matrices:
\begin{eqnarray}\label{P}
   P_{ij;kl} = \langle\Phi_{ij}^P|\Phi_{kl}^P\rangle  = \langle \psi | a_i^{\dagger} a_j^{\dagger}a_la_k|\psi\rangle,~~ P \geq 0 ,
\end{eqnarray}
\begin{eqnarray}\label{Q}
   Q_{ij;kl} = \langle\Phi_{ij}^Q|\Phi_{kl}^Q\rangle  = \langle \psi | a_ia_ja_l^{\dagger} a_k^{\dagger}|\psi\rangle,~~ Q \geq 0,
\end{eqnarray}
\begin{eqnarray}\label{G}
   G_{ik;lj} = \langle\Phi_{ij}^G|\Phi_{kl}^G\rangle  = \langle \psi | a_i^{\dagger}a_k a_j^{\dagger}a_l|\psi\rangle, ~~G \geq 0,
\end{eqnarray}
where $|\Phi_{kl}^P\rangle,|\Phi_{kl}^Q\rangle$, and $|\Phi_{kl}^G\rangle$, are $(N-2)$-, $(N+2)$- , and $N$- particle basis functions, respectively. The $P$ matrix is the 2RDM, and the $Q$ and $G$ matrices can be rewritten as a function of the 1RDM and 2RDM by reordering the creation and annihilation operators using anticommutation relations. One then obtains the following linear maps:
\begin{eqnarray}\label{2Q}
        Q_{ij;kl} &=& \delta_{ik}\delta_{jl} - \delta_{jk}\delta_{il} +P_{ij;kl}  \nonumber \\
        &-& (\delta_{ik}{\rho_{jl}} -\delta_{il}{\rho_{jk}}-\delta_{jk}{\rho_{il}}+\delta_{jl}{\rho_{ik}}), \\
\label{2G}
         G_{ij;kl} &=& \delta_{jl}{\rho_{ik}} - P_{il;kj}.
\end{eqnarray}

The metric matrices associated to the 3-positivity condition read similarly to Eqs. (\ref{P}), (\ref{Q}), and (\ref{G}):
\begin{eqnarray}\label{3D}
         ^3D_{ijk;lmn} = \langle \Psi|a_i^\dagger a_j^\dagger a_k^\dagger a_n a_m a_l | \Psi \rangle,~~^3D \geq 0,
\end{eqnarray}
\begin{eqnarray}\label{3E}
         ^3E_{ijk;lmn} = \langle \Psi|a_i^\dagger a_j^\dagger a_k a_n^\dagger a_m a_l | \Psi \rangle,~~^3E \geq 0,
\end{eqnarray}
\begin{eqnarray}\label{3F}
         ^3F_{ijk;lmn} = \langle \Psi|a_i^\dagger a_j a_k a_n^\dagger a_m^\dagger a_l | \Psi \rangle,~~^3F \geq 0,
\end{eqnarray}
\begin{eqnarray}\label{3Q}
         ^3Q_{ijk;lmn} = \langle \Psi|a_i a_j a_k a_n^\dagger a_m^\dagger a_l^\dagger | \Psi \rangle,~~^3Q \geq 0,
\end{eqnarray}
where $^3D$ and $^3Q$ are the 3p-RDM and 3h-RDM, respectively, and $^3E$ and $^3F$ are  generalizations of the $G$ matrix.  In practice, however, the full 3-positivity condition is too cumbersome to apply. As a consequence, the two less stringent conditions $T_1$ and $T_2$, originally proposed by Erdahl \cite{Erdahl1978}, and implemented for small atoms and molecules by Zhao $et~al.$ \cite{Zhao2004} are used in practical calculations. They read:
\begin{eqnarray}\label{T1}
         (T_1)_{ijk;lmn} = {^3}D_{ijk;lmn} + {^3}Q_{ijk;lmn}, ~~~T_1 \geq 0,
\end{eqnarray}
\begin{eqnarray}\label{T2}
         (T_2)_{ijk;lmn} =  {^3}E_{ijk;lmn} + {^3}F_{ijk;lmn}, ~~~T_2 \geq 0,
\end{eqnarray}
which are both positive semidefinite. The fundamental interest of Eqs. (\ref{T1}) and (\ref{T2}) is that the effected summation of the considered 3-positive metric matrices cause their connected (or cumulant) part to cancel exactly, so that both the $T_1$ and the $T_2$ metric matrices can be rewritten as a function of the 1RDM and 2RDM only:
   \begin{eqnarray}\label{3T1}
   (T_1)_{ijk;lmn} &=&  \delta_{kn}\delta_{jm}\delta_{il} - \delta_{km}\delta_{il}\delta_{jn}+ \delta_{in}\delta_{km}\delta_{jl} \nonumber \\
                 &-&\delta_{kn}\delta_{im}\delta_{jl}+\delta_{jn}\delta_{im}\delta_{kl} -\delta_{in}\delta_{jm}\delta_{kl} \nonumber \\
                  &-&(\delta_{kn}\delta_{jm}-\delta_{jn}\delta_{kn}){\rho_{il}} + (\delta_{kn}\delta_{im}-\delta_{in}\delta_{km}){\rho_{jl}} \nonumber \\
                  &-&(\delta_{jn}\delta_{im}-\delta_{in}\delta_{jm}){\rho_{kl}}+(\delta_{km}\delta_{jl}-\delta_{jn}\delta_{kl}){\rho_{im}}   \nonumber \\
                  &-& (\delta_{kn}\delta_{il}-\delta_{in}\delta_{kl}){\rho_{jm}}+(\delta_{jn}\delta_{il}-\delta_{in}\delta_{jl}){\rho_{km}} \nonumber \\
                  &-& (\delta_{jl}\delta_{km}-\delta_{jm}\delta_{kl}){\rho_{in}}+(\delta_{km}\delta_{il}-\delta_{im}\delta_{kl}){\rho_{jn}} \nonumber \\
                  &-&(\delta_{jm}\delta_{il}-\delta_{im}\delta_{jl}){\rho_{kn}}+\delta_{kn}{P_{ij;lm}}  \nonumber \\
                  &-& \delta_{jn}{P_{ik;lm}} +\delta_{in}{P_{jk;lm}}-\delta_{km}{P_{ij;lm}} \nonumber \\
                  &-& \delta_{jm}{P_{ik;ln}}-\delta_{im}{P_{jk;ln}} +\delta_{kl}{P_{ij;mn}} \nonumber \\
                  &-&\delta_{jl}{P_{ik;mn}}+ \delta_{il}{P_{jk;mn}},
  \end{eqnarray}
\begin{eqnarray}\label{3T2}
      (T_2)_{ijk;lmn}&=& (\delta_{il}\delta_{jm}-\delta_{im}\delta_{jl}){\rho_{kn}} \nonumber \\
     &-& \delta_{kn}{P_{ij;lm}}-\delta_{il}{P_{km;nj}} \nonumber \\
     &-& \delta_{jl}{P_{km;ni}}+\delta_{im}{P_{kl;nj}}-\delta_{jm}{P_{kl;ni}}.
\end{eqnarray}
 The $T_1$ and $T_2$ conditions are, however, inferior to the full 3-positivity condition,
 since they imply the positive semidefinite character of the expectation value of the sum $C_iC_j^\dagger+C_j C_i^\dagger$ with a trial 2RDM, and not that of its individual terms, equal to $C_iC_j^\dagger$ and $C_j C_i^\dagger$.
Usually, an enhancement of the $T_2$ condition, denoted as the the $T_2^\prime$ condition, is used instead of the $T_2$ condition. The $T_2^\prime$ condition arises from the addition of a one-particle operator in the metric defining $T_2$. In the $T_2^\prime$ condition, the positive semidefinite form is equal to:
\begin{eqnarray}\label{3T2+}
      (C_i + B_i)^\dagger (C_j + B_j) + C_jC_i^\dagger,
\end{eqnarray}
where $B_i$ is a one-body operator.  The positive semidefinite character arising from Eq. (\ref{3T2+}) translates into a matrix positivity condition:
\begin{eqnarray}\label{3T2+_positive}
     T_2^\prime =  \left(
 \begin{array}{cc}
     (T_2)_{ijk;lmn} & P_{ij;rk}  \\
     P_{sn;lm}^\dagger & \rho_{sr}
 \end{array}
 \right) \ge 0.
\end{eqnarray}
One can check that the use of $T_2$ or $T_2'$ virtually has the same numerical cost.
For more theoretical and historical details, we refer the reader to Ref. \cite{2RDM_Mazzi}.

Nuclear wave functions are subject to particle number conservation, which can be expressed with the particle and pair number operators $\hat{N}_\nu$ and $\hat{N}_{\nu\nu}$, respectively, reading
\begin{eqnarray}
\hat{N}_\nu &=& \sum_{\alpha_\nu} a^\dagger_{\alpha_\nu} a_{\alpha_\nu} \label{N_op},\\
\hat{N}_{\nu \nu} &=& \sum_{\alpha_\nu < \beta_\nu} a^\dagger_{\alpha_\nu} a^\dagger_{\beta_\nu} a_{\beta_\nu} a_{\alpha_\nu} \label{Npair_op},
 \end{eqnarray}
where $\nu$ represents proton or neutron. Many-body systems obey rotational symmetries as well, i.e.~the total angular momentum operator $J$ and total isospin operator $T$ (in the absence of the Coulomb force for the latter) are conserved quantities. As the expectation values of the operators associated to conservation laws are linear with respect to the 2RDM, they can be directly included as constraints:
\begin{eqnarray}\label{N}
   \langle \Phi | \hat{N}_{\nu\nu} |\Phi \rangle &=&  \text{Tr}~\Gamma_{\nu\nu} = \frac{N_\nu(N_\nu-1)}{2}, \label{Npair} \\
   \langle \Phi | \hat{N}\hat{Z} |\Phi \rangle &=&  \text{Tr}~\Gamma_{pn} = NZ, \label{NZ} \\
   \langle \Phi | \hat{J}^2 |\Phi \rangle &=& \text{Tr}~\hat{J}^2 \Gamma = J(J+1), \label{J2} \\
   \langle \Phi | \hat{T}^2 |\Phi \rangle &=& \text{Tr}~\hat{T}^2 \Gamma  = T(T+1) \label{T2_isospin},
\end{eqnarray}
where $N_\nu = Z$ or $N_\nu = N$,
and the isospin conservation is used only in the absence of the Coulomb force.
Note that proton and neutron number conservation, embodied in Eq. (\ref{particle number}), is a direct consequence of Eqs. (\ref{1RDM}), (\ref{Npair}) and (\ref{NZ}). In particular, this implies that $\langle \Phi | \hat{N}_{\nu}^2 | \Phi \rangle = \langle \Phi | \hat{N}_{\nu} | \Phi \rangle^2$, as $2\hat{N}_{\nu \nu} = \hat{N}_\nu^2 - \hat{N}_\nu$ (see Eqs. (\ref{N_op}) and (\ref{Npair_op})).

Variational 2RDM calculations can be constrained using different conditions. Calculations using the $P$, $Q$ and $G$ 2-positive conditions, combined with the quantum number conservation conditions of Eqs. (\ref{N}), (\ref{NZ}), (\ref{J2}), and (\ref{T2_isospin})  will be denoted by $PQG$. The inclusion of $T_1$ or $T_2^\prime$, or $T_1$ and $T_2^\prime$ will be denoted by $PQG T_1$, $PQG T_2^\prime$, and $PQG T_1 T_2^\prime$, respectively.

Even though Eqs. (\ref{P}), (\ref{Q}), (\ref{G}), (\ref{T1}), and (\ref{3T2+_positive}) make only use of 1RDM and 2RDM, they cannot be used directly in practice. This arises because their computational cost increases quickly with the number of one-body basis states. Indeed, for a number of one-body basis states equal to $s$, the memory needed to store matrices using $PQG$ ($PQG T_1$, $PQG T_2$, $PQG T_1 T_2'$) matrices is $\mathcal{O}(s^4)$ ($\mathcal{O}(s^6)$) while the time of calculation scales as $\mathcal{O}(s^6)$ ($\mathcal{O}(s^9)$), and is independent of the number of valence nucleons. These estimates arise because up to four (six) creation/annihilation operators are present in $PQG$ ($PQG T_1$, $PQG T_2$, $PQG T_1 T_2'$) operators, on the one hand, and because one has to diagonalize the $PQG$ ($PQG T_1$, $PQG T_2$, $PQG T_1 T_2'$) matrices in the constrained energy minimization procedure (see Sec.(\ref{semi_def_opt})), on the other hand, along with a linear system to solve of similar computational cost \cite{Verstichel_PhD}.

Consequently, it is necessary in practice to couple angular momenta in Eqs. (\ref{P}), (\ref{Q}), (\ref{G}), (\ref{T1}), and (\ref{3T2+_positive}). Atomic and molecular systems are considered in the  $LS$ scheme as spin $S$ and orbital $L$ angular momenta can be considered as good quantum numbers \cite{Verstichel_PhD}. Conversely, in nuclear systems, only the total angular momentum $J$ is conserved, so that one uses $jj$ coupling scheme, i.e.,~one works in $J$-scheme. Coupling algebra for that matter is standard, as it makes use of the Wigner-Eckart theorem only, so that the obtained equations are defined from the $J$-coupled 1RDM and 2RDM, Wigner signs and three-body coefficients of fractional parentage \cite{Verstichel_PhD}. The obtained reduction in the demanded memory and time of calculation is huge, as the memory and time of calculation for the $PQG$ ($PQG T_1$, $PQG T_2$, $PQG T_1 T_2'$) case is now $\mathcal{O}(r^4)$ ($\mathcal{O}(r^6)$) and $\mathcal{O}(r^6)$ ($\mathcal{O}(r^9)$), respectively, with $r$ the number of orbitals. Indeed, the ratio $s/r$, which is the average number of states in one basis orbital, is around 5, so that the gain in computational cost can easily reach 10,000-100,000 \cite{Verstichel_PhD}. However, the use of $PQG T_1$, $PQG T_2$ or $PQG T_1 T_2'$ conditions still remains more expensive that of $PQG$ by a large factor, which can reach 100-1000, as one can have tens of orbitals in practical calculations \cite{Verstichel_PhD}. Consequently, it is customary to impose only $PQG$ conditions in atomic and molecular conditions \cite{PhysRevLett.106.083001,Verstichel_PhD}. However, we will see in the following that $T_1$ and $T_2'$ conditions can barely be ignored in nuclear systems (see Sec.(\ref{Results})).

\subsection{Semidefinite optimization} \label{semi_def_opt}

In the variational reduced density matrix theory, the energy of Eq. (\ref{Eg}) is minimized with respect to the 2RDM subject to the constraints outlined above. The solution of the $E_g$ problem is then a semidefinite linear optimization problem with constraints. The primal formulation of this problem, i.e.,~where the 2RDM is considered as a variable to optimize, can be expressed as:
\begin{eqnarray}\label{primal}
   &&   \text{minimize}~~ E_g = \text{Tr}~H\Gamma, \nonumber \\
   &&   \text{such~ that}~~ A\Gamma= X, \nonumber \\
   &&   \text{with} ~~X \geq 0,
\end{eqnarray}
where $\Gamma$ is the primal solution vector and $H$ is the Hamiltonian which contains the one-body and two-body matrix elements. The constraint matrix $A$ and constraint vector $X$ encode the $N$-$representability$ conditions and the equalities of Eqs. (\ref{N}), (\ref{NZ}), (\ref{J2}), and (\ref{T2_isospin}) associated to conserved quantum numbers that the $\Gamma$ matrix must satisfy. The operator $X$ maps the primal solutions onto the set of positive semidefinite RDMs:
\begin{eqnarray}\label{AGamma}
X =
\begin{pmatrix}
~P & 0 & 0 & 0   & 0          & 0 \\
 0 & Q & 0 & 0   & 0          & 0 \\
 0 & 0 & G & 0   & 0          & 0 \\
 0 & 0 & 0 & T_1 & 0          & 0 \\
 0 & 0 & 0 & 0   & T_2^\prime & 0 \\
 0 & 0 & 0 & 0   & 0          & C~
\end{pmatrix}, \label{X}
\end{eqnarray}
where $C$ represents the action of the operators associated to conserved quantum numbers, as it can be written as $C\Gamma = 0$ (see Eqs. (\ref{N}), (\ref{NZ}), (\ref{J2}), and (\ref{T2_isospin})).
In order to find the $\Gamma$ matrix, one will apply the  boundary point positive semidefinite algorithm.
The boundary point method is actually an instance of a more general class of augmented Lagrangian approaches for solving the positive semidefinite problem \cite{PhysRevLett.93.213001,PhysRevLett.106.083001,VERSTICHEL20111235,Verstichel_PhD}.
The standard  Lagrangian for the positive semidefinite problem can be written as,
\begin{eqnarray}\label{La1}
\mathcal{L} = \text{Tr}~H\Gamma  + \Lambda \cdot (X-A\Gamma),
\end{eqnarray}
to minimize the $H\Gamma$ functional and where a Lagrangian multiplier matrix $\Lambda$ is introduced to account for constraints. While Eq. (\ref{La1}) would provide the $\Gamma$ matrix from a theoretical point of view, it cannot be used in practice because the convergence would be very poor, or even nonexistent numerically \cite{ALM}. This arises because the functional of Eq. (\ref{La1}) is not convex, i.e.,~it does not resemble a paraboloid in the vicinity of the sought minimum \cite{ALM}.

In order to account for this problem, the augmented Lagrangian method for primal problems has been developed. For this, one adds a quadratic penalty for the so-called infeasibility, i.e.,~the fact that $\Gamma$ does not belong to the set of acceptable solutions during optimization because $A\Gamma \neq X$ \cite{Zhao2004}:
\begin{eqnarray}\label{La2}
\mathcal{L} = \text{Tr}~H\Gamma  + \Lambda \cdot (X-A\Gamma) + \frac{\sigma}{2} || X-A\Gamma||^2,
\end{eqnarray}
where a penalty parameter, $\sigma > 0$, is introduced and determines how stringently the  $N$-$representability$ constraint is enforced.  As Eq. (\ref{La2}) is quadratic in $\Gamma$, its minimum is straightforward to find with standard minimization methods if $X$ is kept fixed \cite{Verstichel_PhD}. $X$ is then obtained as a second step also from the minimization of Eq. (\ref{La2}), but by keeping $\Gamma$ fixed \cite{Verstichel_PhD}. In particular, $X$ positivity is enforced by way of a diagonalization procedure and matrix manipulations, where the negative eigenvalues of the $P$, $Q$, $G$, $T_1$, $T_2'$ metric matrices and the non-zero eigenvalues of the $C$ matrix in Eq. (\ref{X}) are replaced by zero \cite{Verstichel_PhD,PhysRevLett.106.083001}.
$\sigma$ is modified at each iteration according to the values of $\nabla \mathcal{L}$ and $|| X-A\Gamma||$ to accelerate convergence \cite{Burer_Monteiro,Verstichel_PhD}.
One then obtains an iterative procedure, where, at each iteration, $\Gamma$ is the minimum of a slightly different problem as that of Eq. (\ref{primal}) with $||A\Gamma - X|| \rightarrow 0$ at convergence (see Refs.\cite{Verstichel_PhD,PhysRevLett.106.083001} for details).

The boundary point method has been successfully used in the variational 2RDM calculations for atomic systems, as the calculated ground-state energies agree well with those obtained from those implemented from full configuration interaction approaches \cite{PhysRevLett.106.083001,PhysRevLett.108.213001,PhysRevLett.99.243002}.

The number of iterations to reach convergence when using the augmented Lagrangian method is about one to a few thousands for atomic and molecular systems \cite{Verstichel_PhD,PhysRevLett.106.083001}. The situation is similar when applying this method to nuclei according to our numerical tests. However, the number of iterations can easily reach 10,000 due to the rather large strength of the nucleon-nucleon interaction. Added to that, it can happen that the augmented Lagrangian method does not converge when $N$-$representability$ constraints are not sufficiently strong (see Sec.(\ref{Results})).

Because the variational 2RDM conditions are independent of a  reference wave function, they can treat both moderate and strong coupling systems, such as atomic and nuclear systems.   Furthermore, closed-shell and open-shell systems, as well as nuclei with an even or odd number of protons and neutrons, are treated equally in the variational 2RDM calculations, so that it is as general as configuration interaction methods from that point of view. This is in contrast with coupled-cluster theory, for example, where a reference state is used and only closed-shell systems plus or minus one to two nucleons can be used \cite{hagen2016emergent}.

\section{Results} \label{Results}

\subsection{Variational 2RDM calculations with cluster-orbital shell-model Hamiltonian} \label{COSM_RDM}

The first example will be considered with a phenomenological Hamiltonian for simplicity, where one uses the core + valence particle picture.
We will then apply the variational 2RDM method using the cluster-orbital shell-model  (COSM) Hamiltonian, which is the natural framework when using core and valence nucleons with an effective Hamiltonian \cite{Suzuki1988}. The COSM coordinates of the valence nucleons are defined from the center of the core nucleus, and the kinetic energy of the center of mass motion is subtracted from the total $A$-body Hamiltonian. As a consequence, no center of mass motion can arise in COSM. In fact, the COSM coordinates form a set of Jacobi coordinates.
The Hamiltonian in the COSM frame reads:
\begin{eqnarray}\label{COSM_H}
\hat{H} = \sum_{i=1}^{N_{\text{val}}}  \left (\frac{\hat{\mathbf{p}}_i^2}{2\mu_i} + \hat{U}^{(c)}_i \right)+ \sum_{i<j}^{N_{\text{val}}} \left ( \hat{V}_{ij} + \frac{\hat{\mathbf{p}}_i \cdot \hat{\mathbf{p}}_j}{M_c} \right),
\end{eqnarray}
where $N_{\text{val}}$ is the number of valence nucleons, $\mu_i$ is the reduced mass of the $i$-th nucleon with respect to the core, and $M_c$ is the mass of the core. The single-particle potential $\hat{U}^{(c)}_i $ mimics the field generated by the core on each nucleon.  It is represented by a Woods-Saxon (WS) potential with a spin-orbit term. $\hat{V}_{ij}$ is the two-body residual interaction, which is composed of the nucleon-nucleon interaction and the two-body Coulomb interaction. The additional two-body kinetic term in Eq. (\ref{COSM_H}), depending on $M_c$, is the recoil term of the COSM Hamiltonian and arises from the translationally invariant character of the COSM framework.

As we aim at comparing the numerical results arising from the variational 2RDM method in nuclear systems to exact eigenenergies, it is sufficient to consider a simple nucleon-nucleon interaction for $V_{ij}$. Thus, the \textit{surface~Gaussian~interaction} (SGI) \cite{PhysRevC.89.034624} is used :
\begin{eqnarray}\label{SDI}
&& V_{\text{SGI}}(\mathbf{r}_i-\mathbf{r}_j) \nonumber \\
&& =  V_0~ \text{exp} \left [ -\left (\frac{\mathbf{r}_i-\mathbf{r}_j}{\mu_I}\right )^2 \right ]\delta(r_i-r_j-2R_0),
\end{eqnarray}
where $\mu_I$ is the interaction range; $V_0$ is the strength of the interaction and $R_0$ is the nuclear radius. In the present work, we fix the interaction range and nucleus radius to $\mu_I$ = 1 fm and $R_0 = 2$ fm, respectively. $V_0$ is adjusted to reproduce the selected experimental data.

As a first application of the variational 2RDM method, we will compare the variational eigenenergies of $^{6,8}$He ground states obtained in the variational 2RDM method with shell-model results, which can be deemed as exact for that matter. $^4$He is selected as the inner core, and the parameters of the one-body WS potential mimicking the core are taken from Ref.\cite{PhysRevC.96.054316}. Calculations are performed in a model space spanned by the $sp$ partial waves. For each partial wave, six harmonic oscillator (HO) shells are used. The model space is consists of 17 orbitals, excluding the core space. The  $V_0$ coupling constant of the SDI interaction is adjusted to reproduce the binding energy of $^{6}$He \cite{ensdf}. Its strength is equal to $-435$ MeV.

\begin{table}[!htb]
\centering
\caption{\label{RDM_6_8_He} Ground-state energies of $^{6,8}$He (in MeV), calculated with SM and variational 2RDM methods using the COSM Hamiltonian. Variational 2RDM calculations are constrained with $PQG$ conditions. Experimental data \cite{ensdf} are also provided.}
\setlength{\tabcolsep}{2.96mm}{
\begin{tabular}{lcccc} \hline \hline
Nuclei        & $PQG$        & SM       & Exp \\  \hline
 $^6$He        & $-1.090$ &  $-0.974$   & $-0.975$  \\
 $^8$He        & $-5.845$ &  $-5.795$  &  $-3.289$ \\  \hline \hline
\end{tabular}}
\end{table}

\begin{figure}[!htb]
\includegraphics[width=1.00\columnwidth]{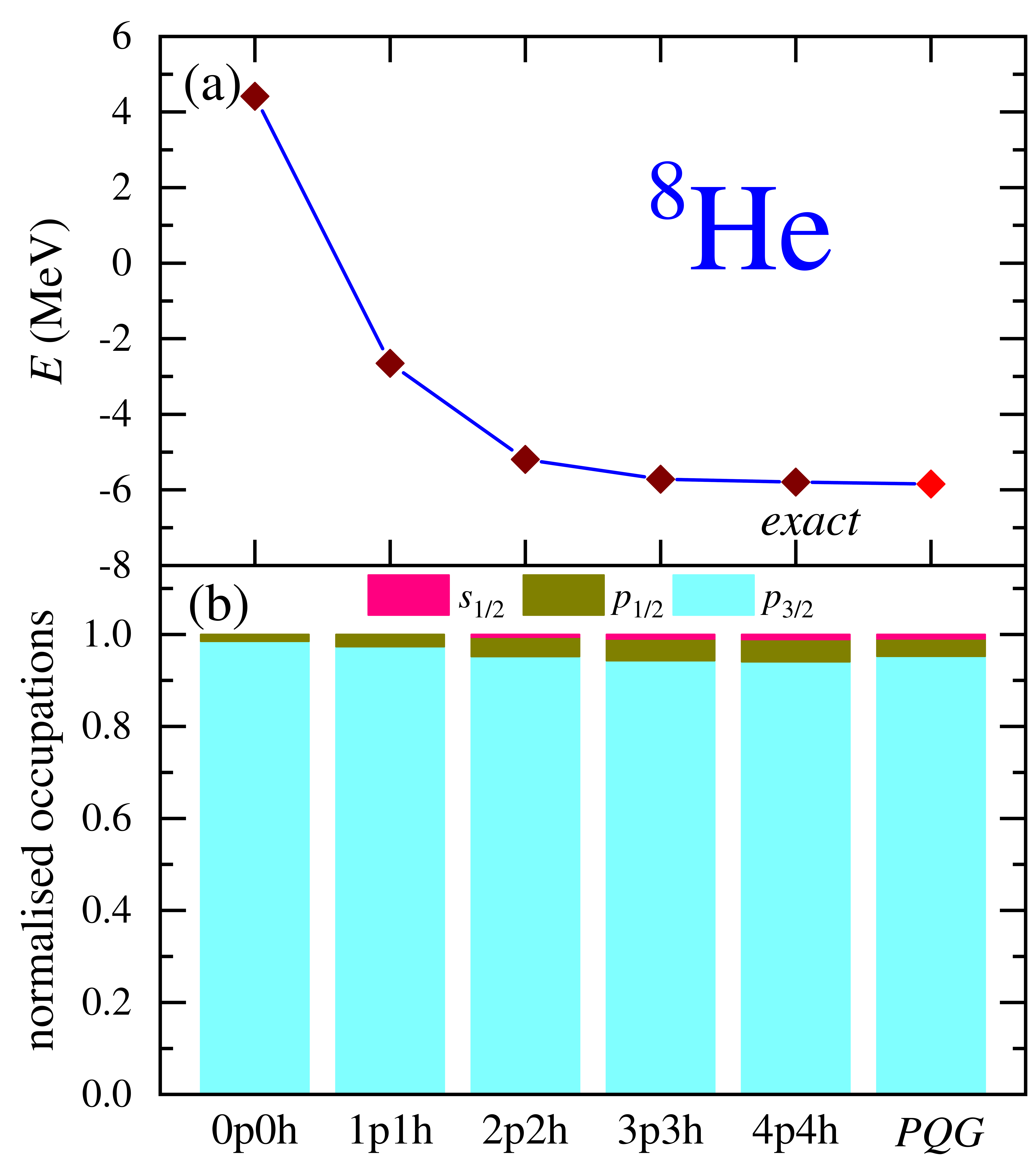}
\caption{Energy and normalized occupations of $^8$He ground state using SM and variational 2RDM calculations. COSM calculations (brown diamond) are performed using different truncation schemes from 0p0h to 4p4h (exact). Variational 2RDM calculations (red diamond) are constrained with $PQG$ conditions.}{\label{8He}}
\end{figure}

Results obtained with variational 2RDM and shell-model are shown in Tab. \ref{RDM_6_8_He}, along with experimental data \cite{ensdf}.
Due to the small model space used, the obtained $^{8}$He ground-state energy is overbound in our calculations when compared with experimental data \cite{ensdf}. However, this is not important for our purpose as we only aim at benchmarking the variational 2RDM results with exact energies.  We can see that the variational 2RDM calculations of $^{6,8}$He are close to the exact COSM results, where the largest difference is about 100 keV.

Let us now compare partial wave occupation in variational 2RDM and COSM frameworks for $^8$He. For this, one performs several COSM calculations in truncated spaces, where one allows 0, 1, $\dots$, 4 neutrons in orbitals outside the $0p$-shell. These truncated spaces are labeled by 0p0h, $\dots$, 4p4h, respectively.
Normalized occupation of the $sp$ partial waves are shown in Fig. \ref{8He}(b).  Normalized occupations are obtained by being divided by the number of valence particles. As can be expected, the occupation of $s_{1/2}$ and $p_{1/2}$ partial waves increases with the size of the model space, as they are generated by the configuration mixing induced by the nucleon-nucleon interaction. Occupations of $sp$ partial waves in the variational 2RDM and COSM methods become close in 3p3h and 4p4h. At the same time, the ground-state energy of $^8$He is almost the same in both methods. This means that the 2RDM calculation of the ground state obtained with $PQG$ conditions is close to the exact 2RDM. Hence, in this case, the variational 2RDM allows to calculate almost exactly the ground-state energy of $^8$He, even when the simplest variational 2RDM framework is used, where only the \textit{N-representability} $PQG$ conditions are constrained.

\begin{figure}[!htb]
\includegraphics[width=1.00\columnwidth]{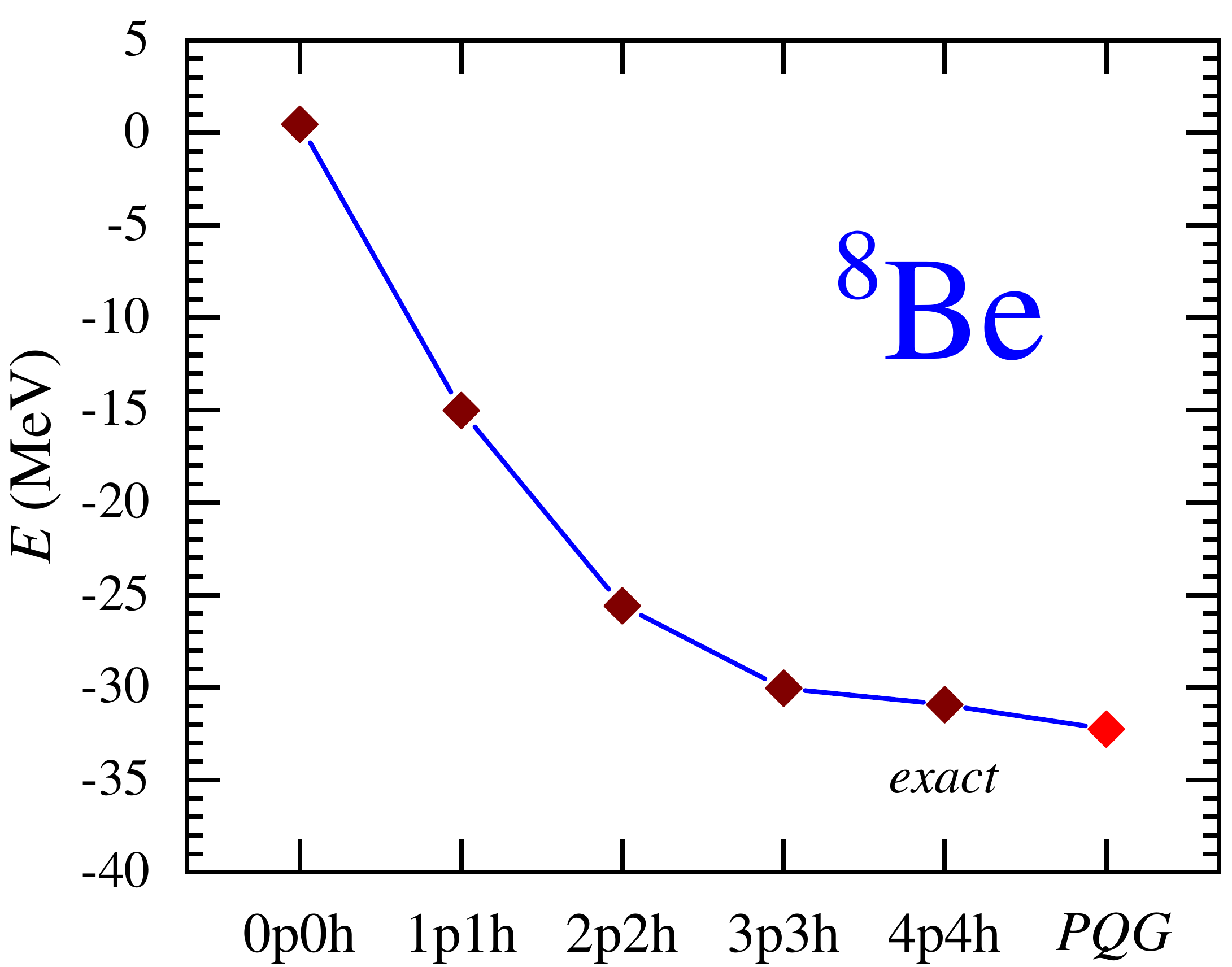}
\caption{Similar to Fig. \ref{8He}(a), but for $^8$Be.}{\label{8Be_E}}
\end{figure}

\begin{figure}[!htb]
\includegraphics[width=1.00\columnwidth]{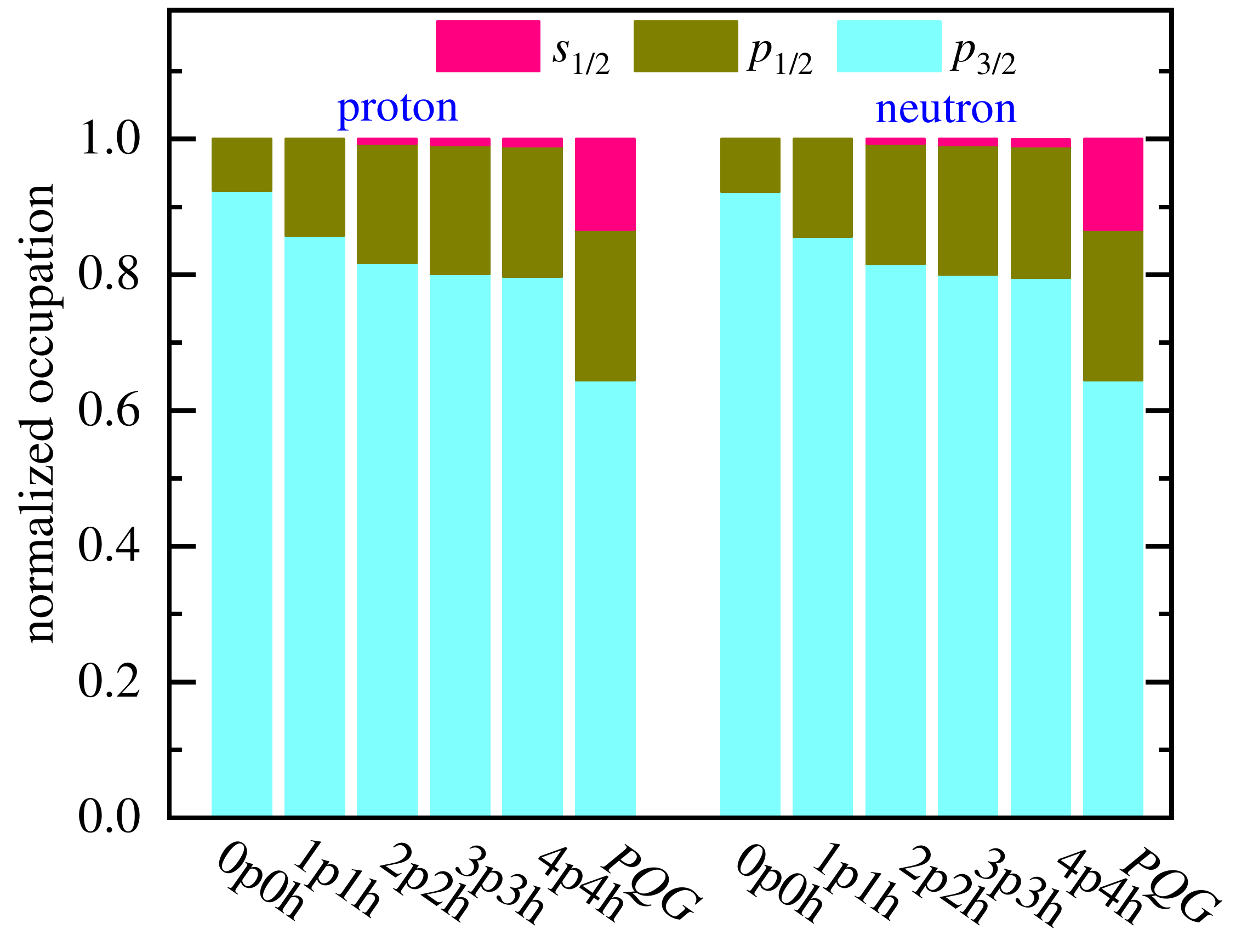}
\caption{Similar to Fig. \ref{8He}(b), but for proton and neutron occupations in $^8$Be.}{\label{8Be_occupation}}
\end{figure}

In the calculations of $^{6,8}$He ground states, only neutrons are considered in the valence space. Consequently, it is interesting to consider a nucleus in which both valence protons and neutrons are present. Thus, we performed variational 2RDM calculations for the $^8$Be ground state, where two valence protons and two valence neutrons are present outside the $^4$He core. The same COSM Hamiltonian as for $^{6,8}$He is used for $^8$Be, except that the Coulomb force is added in proton space. The model space used for protons is also the same as that of neutrons. The ground state of $^8$Be obtained in  the variational 2RDM with $PQG$ conditions are presented in Fig. \ref{8Be_E}, along with COSM calculations with different truncations.
We can see that the ground-state energy of $^8$Be calculated using variational 2RDM with $PQG$ condition is close to the exact results. However, the calculated normalized occupations (see Fig. \ref{8Be_occupation}) show that large differences occur between COSM and variational 2RDM calculations. A large percentage of $s_{1/2}$ partial wave occupation occurs using variational 2RDM, of about 12\%, while it is only about 2\% in the exact wave function provided by diagonalization within COSM. Consequently, the occupation of $p_{3/2}$ partial waves is reduced due to the large $s_{1/2}$ component in the 2RDM calculation, while that of the $p_{1/2}$ partial is almost the same in both variational RDM and COSM calculations. This discrepancy translates into an energy overbinding of the $^8$Be ground state using the variational 2RDM method (see Fig. \ref{8Be_E}). Indeed, the $^8$Be ground state obtained with the variational 2RDM method is too bound by 1.7 MeV compared with the COSM calculation. Thus, this energy overbinding is generated by spurious couplings mainly involving the $s_{1/2}$ and $p_{3/2}$ partial waves.

Hence, in order to obtain precise partial wave occupations of the $^8$Be ground state using the variational 2RDM, it is necessary to include higher-order conditions  \cite{Erdahl1978,Zhao2004} or Hamiltonian-based constraints, i.e.,~inequalities involving the expectation value of two-body operators deduced from the variational principle  \cite{PhysRevLett.105.213003}.
However, the computational cost of the variational 2RDM induced by the $T_1$ and $T_2^\prime$ conditions is expensive when more than a few basis orbitals are present (see Sec.(\ref{N_rep})), so that we could not include these constraints in the variational 2RDM calculations using the COSM Hamiltonian.

The variational 2RDM calculations with $PQG$ condition could provide good descriptions of $^{6,8}$He ground-state energies. The partial wave occupations obtained with the variational 2RDM method are close to those arising from COSM calculations using many-body wave functions. For $^8$Be, the obtained energy with the variational 2RDM method is close to the exact value. However, the partial wave occupations obtained therein show discrepancies when compared to exact values. 

\subsection{Variational 2RDM calculations with USDB shell-model interactions} \label{USDB_RDM}

The variational 2RDM calculations with $PQG$ conditions can provide the ground-state energies of $^{6,8}$He and $^8$Be using a COSM Hamiltonian within a model space built from $sp$ partial waves.
However, the partial wave occupations obtained for $^8$Be are different from the exact values by about 10 \%. This discrepancy might arise from the fact that one has both valence protons and neutrons or because we only constrained the variational 2RDM calculation using $PQG$ conditions. Hence, one would like to be able to apply $T_1$ and $T_2'$ conditions as well. For this, one can only have a few basis orbitals, as otherwise calculations become too expensive numerically. Consequently, we will study the variational 2RDM method using
the USDB interaction \cite{PhysRevC.74.034315} as a testing ground.
The USDB interaction is a standard shell-model interaction for $sd$-shell nuclei, which can provide good descriptions  of the nuclei, such as their the binding energies and  spectra \cite{PhysRevC.74.034315}.  The model space for valence protons and neutrons consists of the $0d_{5/2}$, $1s_{1/2}$ and $0d_{3/2}$ shells in the USDB interaction, so that the small number of basis orbitals makes this framework convenient to study the effect of $T_1$ and $T_2'$ conditions.

For the USDB interaction,  the Hamiltonian is written as
\begin{eqnarray}\label{USDB_H}
\hat{H} = \hat{H}_0 + \hat{H}_1,
\end{eqnarray}
where $\hat{H}_0$ represents the single-particle energies of orbitals in the model space, $\hat{H}_1$ is the residual two-body interaction.
Isospin-breaking effects are not considered in the USDB interaction \cite{PhysRevC.74.034315}.
In the present work, we will first consider the ground-state energies of even-even oxygen isotopes, in which the nuclei only have valence neutrons; after that, we will focus on the $N = Z$ even-even nuclei, thus bearing both valence protons and neutrons, which consist of $^{20}$Ne, $^{24}$Mg, $^{28}$Si, and $^{32}$S.
Due to the isospin conserving character of the USDB interaction, the configurations of valence neutrons and protons are the same in $N = Z$ nuclei. In the latter results, we then just show the normalized occupations of valence neutrons.

\begin{table}[!htb]
\centering
\caption{\label{O_USDB} Ground-state energies of oxygen isotopes, calculated with the variational 2RDM method using the USDB interaction. Variational 2RDM calculations are constrained with $PQG$, $PQG T_1$, $PQG T_2^\prime$, or $PQG T_1 T_2^\prime$ conditions. The star indicates that the variational 2RDM does not converge using the considered constraints. Results are compared with SM calculations.}
\setlength{\tabcolsep}{2.06mm}{
\begin{tabular}{lccccc} \hline \hline
Nuclei        &  $^{20}$O    & $^{22}$O & $^{24}$O & $^{26}$O   \\  \hline
 $PQG$          & $-30.344$ &  $-58.055$   & $*$ & *  \\
 $PQG T_1$        & $-29.701$ &  $-36.404$  &  $ -42.317$ & * \\
 $PQG T_2^\prime$        & $-23.712$ &  $-34.554$  &  $-41.257$ & $-40.869$ \\
 $PQG T_1 T_2^\prime$       & $-23.712$ &  $-34.554$  &  $-41.256$ & $-40.866$ \\
 SM           & $-23.632$ &  $-34.498$  &  $-41.225$ & $-40.868$ \\  \hline\hline
\end{tabular}}
\end{table}

\begin{figure}[!htb]
\includegraphics[width=1.00\columnwidth]{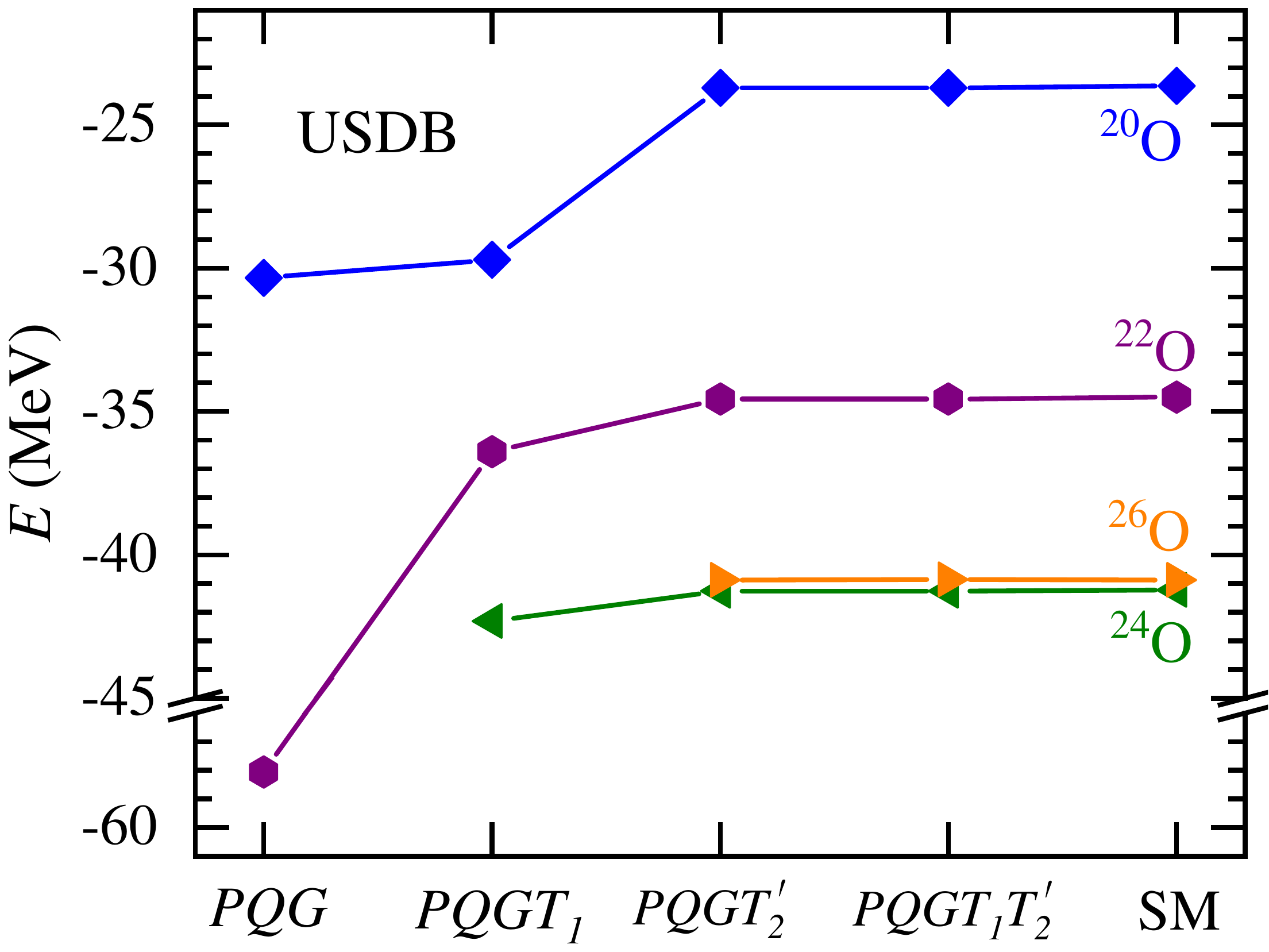}
\caption{Variational 2RDM calculations of oxygen isotopes with different \textit{N-representability} conditions, i.e.,~ $PQG$, $PQG T_1$, $PQG T_2^\prime$, and $PQG T_1 T_2^\prime$ constraints. Results are compared with SM calculations.}{\label{USDB_oxygen_E}}
\end{figure}

\begin{figure}[!htb]
\includegraphics[width=1.00\columnwidth]{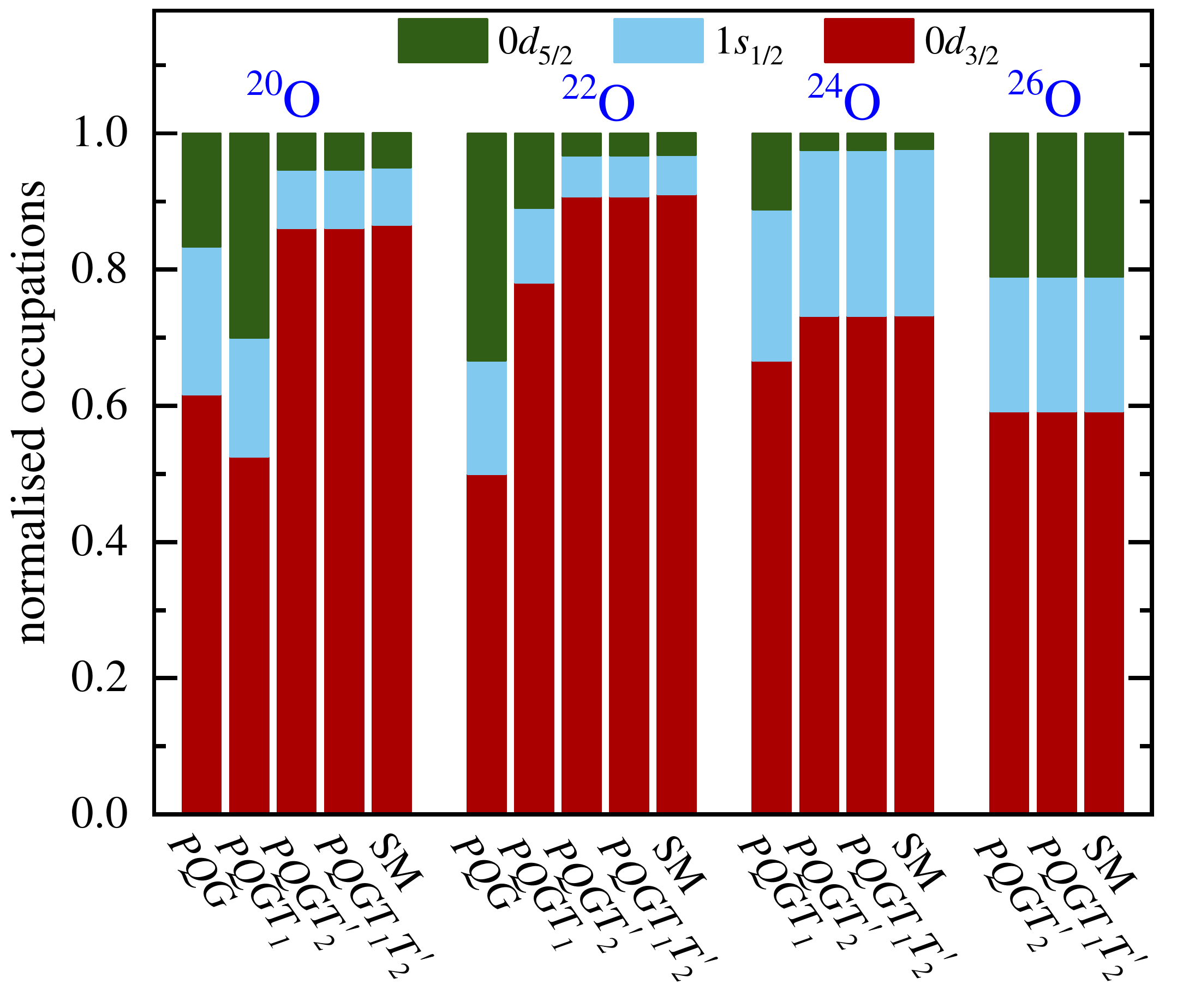}
\caption{Calculated normalized occupations of the ground-state energies of oxygen isotopes using variational 2RDM, with different \textit{N-representability} conditions, and SM.}{\label{USDB_Oxygen_occupation}}
\end{figure}

The calculated ground-state energies of oxygen isotopes using variational 2RDM with different \textit{N-representability} conditions are presented in Tab. \ref{O_USDB} and  Fig. \ref{USDB_oxygen_E}. The results are also compared with the exact values computed from a SM calculation.  We focus on the $^{20\text{-}26}$O even-even isotopes in the oxygen chain.  For $^{24,26}$O, the variational 2RDM calculations using $PQG$ and/or $PQG T_1$ conditions are not shown because one could not obtain convergence of the variational 2RDM method. From Fig. \ref{USDB_oxygen_E}  and Tab. \ref{O_USDB}, we can see that the results using variational 2RDM calculations with $PQG$ conditions are overbound when compared with the exact results. For example, the ground-state energy of $^{22}$O is overbound by about 25 MeV, which is excessively large. The situation can be slightly improved by including the $T_1$ condition in $PQG$, but the results of variational 2RDM calculations with $PQG T_1$ are still overbound by a few MeV.
The energy overbinding is fully overcome in the variational 2RDM by including the $T_2^\prime$ condition. The results of the variational 2RDM with $PQG T_2^\prime$ and $PQG T_1 T_2^\prime$ are indeed very close to the exact results, with differences within 100 keV. One can also note that the results obtained using the $PQG T_1 T_2^\prime$ and $PQG T_2^\prime$ conditions are nearly the same. This is consistent with the conclusions obtained in the study of molecular systems \cite{PhysRevA.72.032510,PhysRevA.74.032501,Verstichel_PhD}.

Let us now consider partial-wave occupations (see Fig. \ref{USDB_Oxygen_occupation}). The variational 2RDM calculations obtained with $PQG T_2^\prime$ or $PQG T_1 T_2^\prime$ are close to the  exact results. However, the calculations with $PQG$ or $PQG T_1$ do not give accurate calculations.
In Tab. \ref{O_USDB}, the detailed values of the obtained ground-state energies are presented. We can see that the largest difference is about 80 keV (which happens in $^{20}$O) when  the variational 2RDM calculations using $PQG T_1 T_2^\prime$ condition are compared with the exact shell-model results.
For $^{26}$O, the variational 2RDM calculation with $PQG T_1 T_2^\prime$ conditions is slightly higher than the exact results, which may be caused by the computational error in the variational calculations.

\begin{figure}[!htb]
\includegraphics[width=1.00\columnwidth]{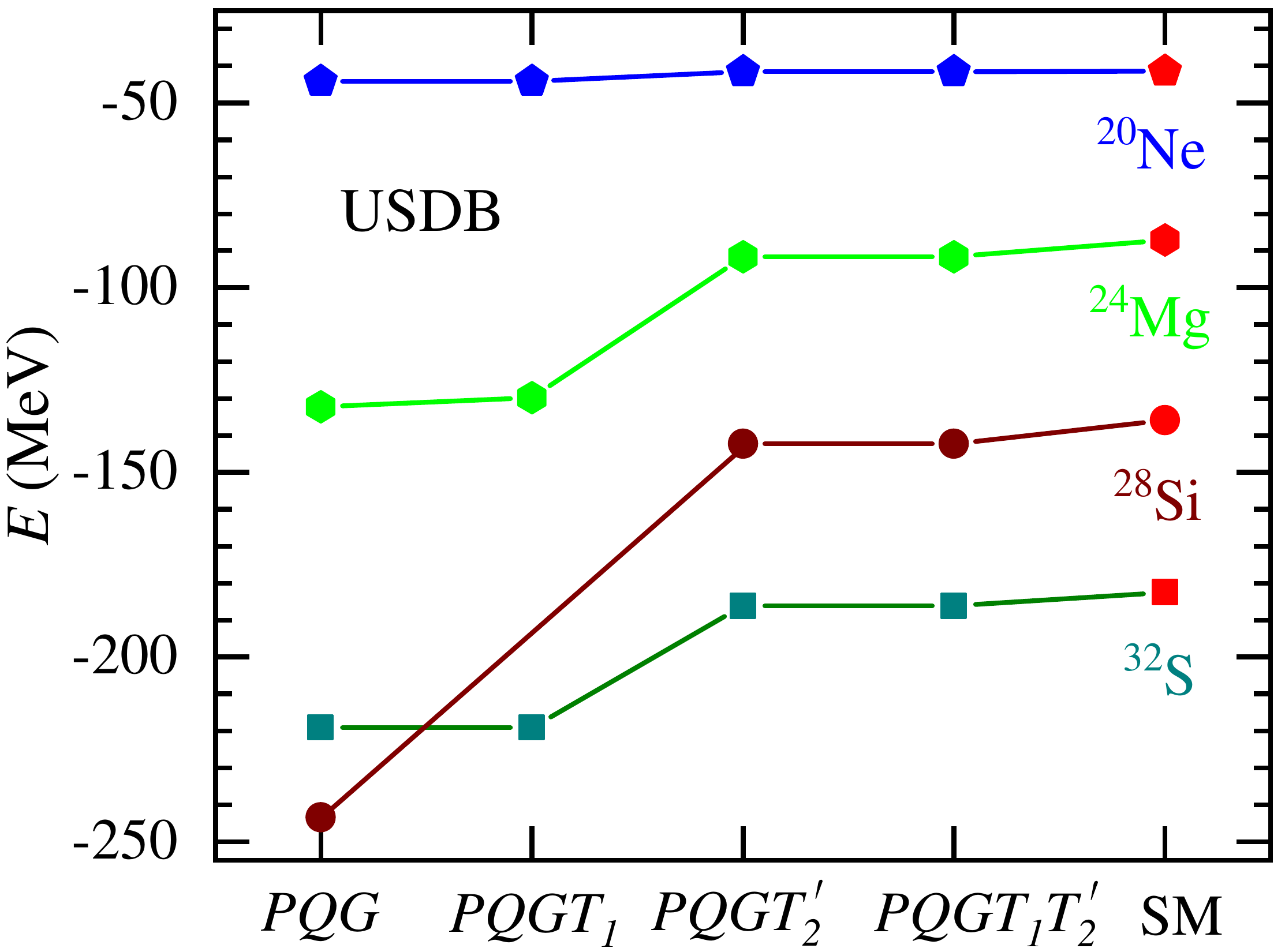}
\caption{Similar to Fig. \ref{USDB_oxygen_E}, but for the $N=Z$ even-even nuclei, $^{20}$Ne, $^{24}$Mg, $^{28}$Si, and $^{32}$S.}{\label{sd_USDB_RDM}}
\end{figure}

\begin{figure}[!htb]
\includegraphics[width=1.00\columnwidth]{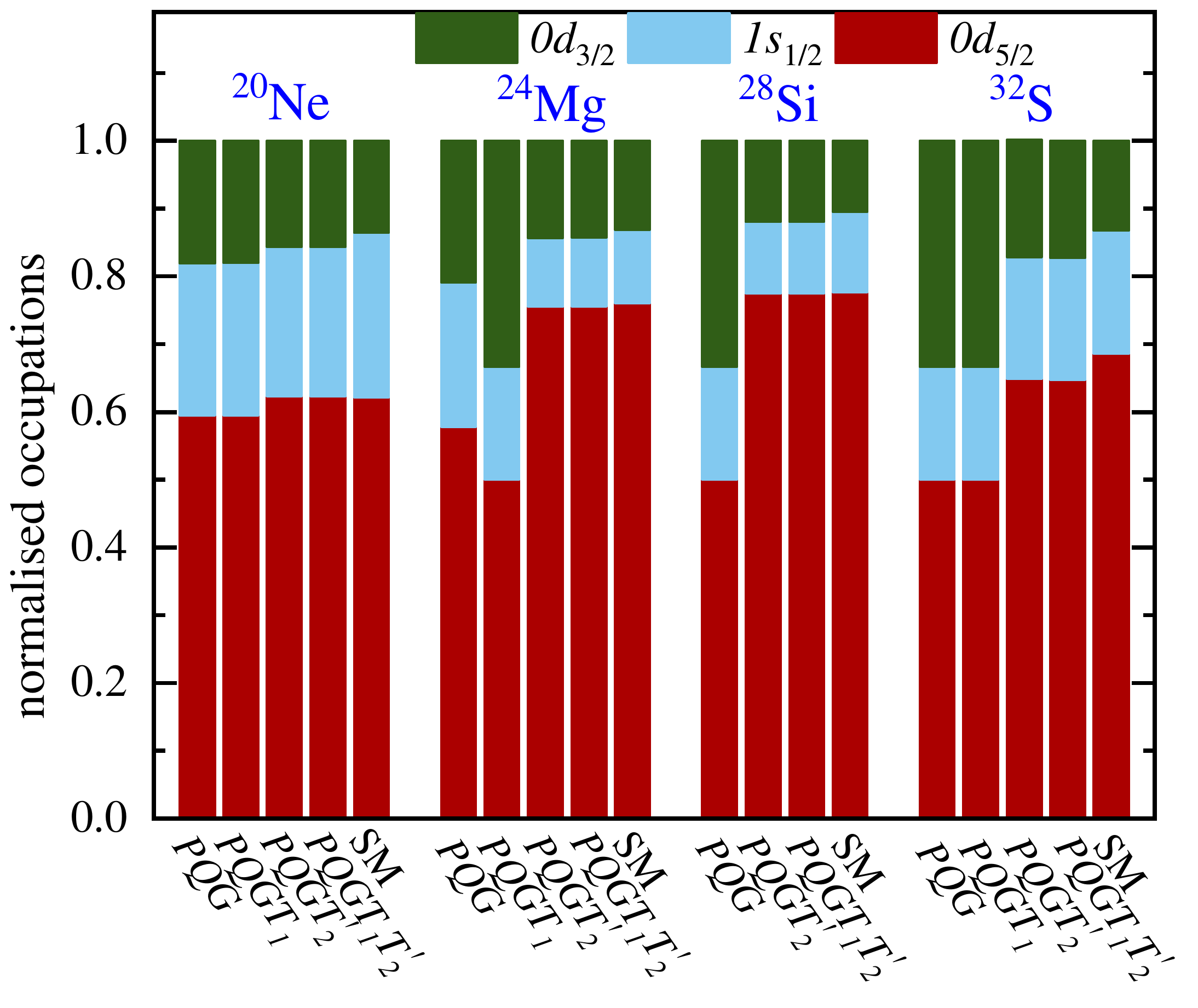}
\caption{Similar to Fig. \ref{USDB_Oxygen_occupation}, but for the $N=Z$ even-even nuclei $^{20}$Ne, $^{24}$Mg, $^{28}$Si, and $^{32}$S.}{\label{USDB_occupation}}
\end{figure}

In order to ponder out the influence of the isospin $T=0$ part of the nucleon-nucleon interaction, we now consider nuclei in which protons and neutrons are both active.  As testing grounds for our benchmarking purpose, we calculate the ground states of the $N = Z$ even-even $sd$-nuclei, $^{20}$Ne, $^{24}$Mg, $^{28}$Si, and $^{32}$S. The results are shown in Tab. \ref{sd_USDB} and Fig. \ref{sd_USDB_RDM}.
Similar to oxygen isotopes, we see that the variational 2RDM calculations using $PQG$ conditions are too overbound. The situation is even worse when one has both valence protons and neutrons, because overbinding can reach 50-100 MeV. This is clearly due to the presence of $T=0$ matrix elements in the nucleon-nucleon interaction, which are about twice larger than $T=1$ matrix elements. However, the situation is not improved when including the $T_1$ condition along with $PQG$ conditions. In fact, only the consideration of the $T_2^\prime$ constraint could significantly ameliorate the quality of calculations. Note that the variational 2RDM calculations with $PQG T_1 T_2^\prime$ constraints are similar to those obtained with $PQG T_2^\prime$ conditions.
However, the discrepancies between variational 2RDM using $PQG T_1 T_2^\prime$ and exact results using SM are still large. The largest energy difference occurs in $^{28}$Si, where it is about 6.5 MeV. Conversely, the energy discrepancy is small in $^{20}$Ne, as it is only about 150 keV.
Normalized partial-wave occupations are presented in Fig. \ref{USDB_occupation}. The variational 2RDM  calculations with $PQG$ and $PQG T_1$ conditions give similar results, as both of them show large discrepancies when compared with exact results.
Partial-wave occupations are largely improved by including the $T_2^\prime$ condition.
The variational 2RDM calculations using $PQG T_2^\prime$ and $PQG T_1 T_2^\prime$  provide nearly the same occupations. The results are close to SM calculations, but the variational 2RDM calculations, including the $T_2^\prime$ condition, still show small discrepancies. This explains ground-state energy overbinding (See Fig. \ref{sd_USDB_RDM}).

\begin{table}[!htb]
\centering
\caption{\label{sd_USDB} Similar to Tab. \ref{O_USDB}, but for the $N=Z$ even-even nuclei $^{20}$Ne, $^{24}$Mg, $^{28}$Si, and $^{32}$S.}
\setlength{\tabcolsep}{2.06mm}{
\begin{tabular}{lccccc} \hline \hline
Nuclei                 &  $^{20}$Ne &  $^{24}$Mg  & $^{28}$Si  & $^{32}$S   \\  \hline
 $PQG$                 & $-44.160$  &  $-129.809$ & $-243.363$ & $-219.061$ \\
 $PQG T_1$             & $-44.167$  &  $-129.808$ & $*$        & $-219.058$ \\
 $PQG T_2^\prime$      & $-41.536$  &  $-91.656$  & $-142.332$ & $-186.186$ \\
 $PQG T_1 T_2^\prime$  & $-41.537$  &  $-91.672$  & $-142.314$ & $-186.314$ \\
 SM                    & $-41.397$  &  $-87.104$  & $-135.861$ & $-182.452$ \\  \hline\hline
\end{tabular}}
\end{table}

The variational 2RDM calculations using USDB Hamiltonian have shown that the $PQG T_1 T_2^\prime$ conditions tremendously improve the precision of calculations. While $PQG$ conditions can lead to an energy overbinding which may even reach 100 MeV, it is decreased by more than an order of magnitude with $PQG T_1 T_2^\prime$ conditions. However, a precision of 5-10 MeV is still not sufficient for nuclear binding energy. Stronger or additional physical conditions \cite{PhysRevLett.108.263002} should be included to improve the calculations of those nuclei.

Nevertheless,  $PQG$ conditions can provide good descriptions for atomic or molecular systems \cite{PhysRevA.71.062503}, where numerical precision is comparable to that of coupled-cluster calculations with singles and doubles excitations \cite{PhysRevA.74.032501,PhysRevA.71.062503}. The ground states of $^{6,8}$He can also be well reproduced using only $PQG$ constraints with a COSM Hamiltonian within an $sp$ model space.

The crux of the problem seems to be how soft the inter-particle interaction is.
In atomic or molecular systems, one only has an interaction of the Coulomb type, whose induced inter-particle correlations are weak. Indeed, the Hartree-Fock approximation provides with more than 99\% of the exact wave function, while electron-electron correlation energy is typically 0.1\% of the total binding energy \cite{PhysRevLett.117.153001}. This is in sharp contrast with the nucleon-nucleon interaction, which provides correlations so strong that they can generate nuclear deformation and clustering \cite{RevModPhys.90.035004}. In fact, mean-field and residual interaction have comparable effects on  nuclear many-body wave functions, as SM calculations have shown \cite{RevModPhys.77.427}.

\begin{table}[!htb]
\centering
\caption{\label{USDB_half_Table} Similar to Tab. \ref{sd_USDB}, but using the USDB(1/2) interaction in the Hamiltonian.}
\setlength{\tabcolsep}{2.06mm}{
\begin{tabular}{lccccc} \hline \hline
Nuclei        &  $^{20}$Ne    & $^{24}$Mg & $^{28}$Si & $^{32}$S   \\  \hline
 $PQG$        & $-28.098$ &  $-77.242$   & $-132.453$ & $-130.003$  \\
 $PQG T_1$        & $-28.089$ &  $-72.090$  &  $-96.748$ & $-124.090$ \\
 $PQG T_2^\prime$        & $-27.150$ &  $-58.852$  &  $-90.886$ & $-119.759$ \\
 $PQG T_1 T_2^\prime$       & $-27.011$ &  $-58.848$  &  $-90.885$ & $-119.756$ \\
 SM           & $-26.627$ &  $-57.007$  &  $-89.294$ & $-119.298$ \\  \hline\hline
\end{tabular}}
\end{table}

\begin{figure}[!htb]
\includegraphics[width=1.00\columnwidth]{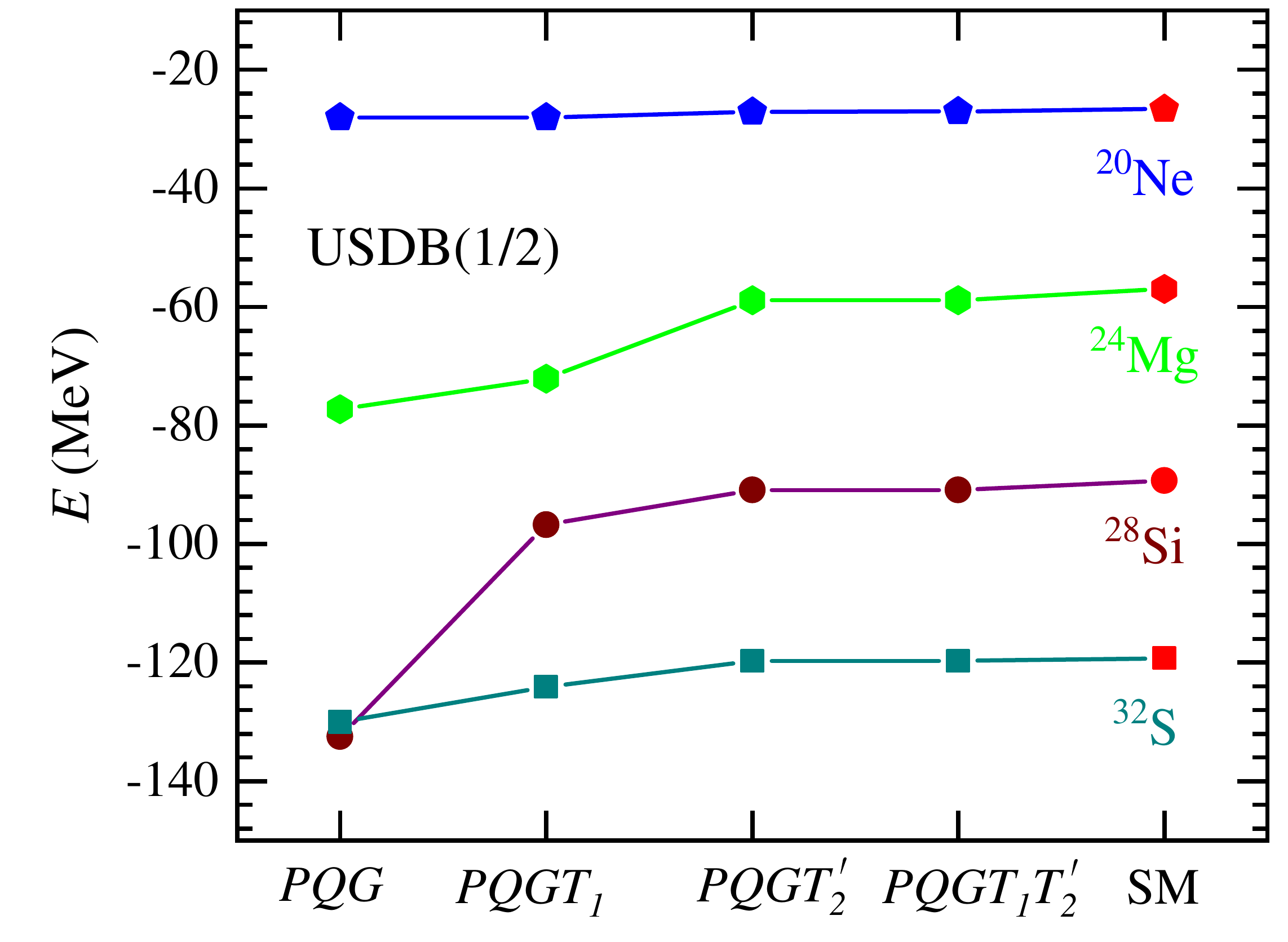}
\caption{Similar to Fig. \ref{sd_USDB_RDM}, but where the USDB(1/2) interaction is utilized as Hamiltonian.}{\label{USDB_half_E}}
\end{figure}

\begin{figure}[!htb]
\includegraphics[width=1.00\columnwidth]{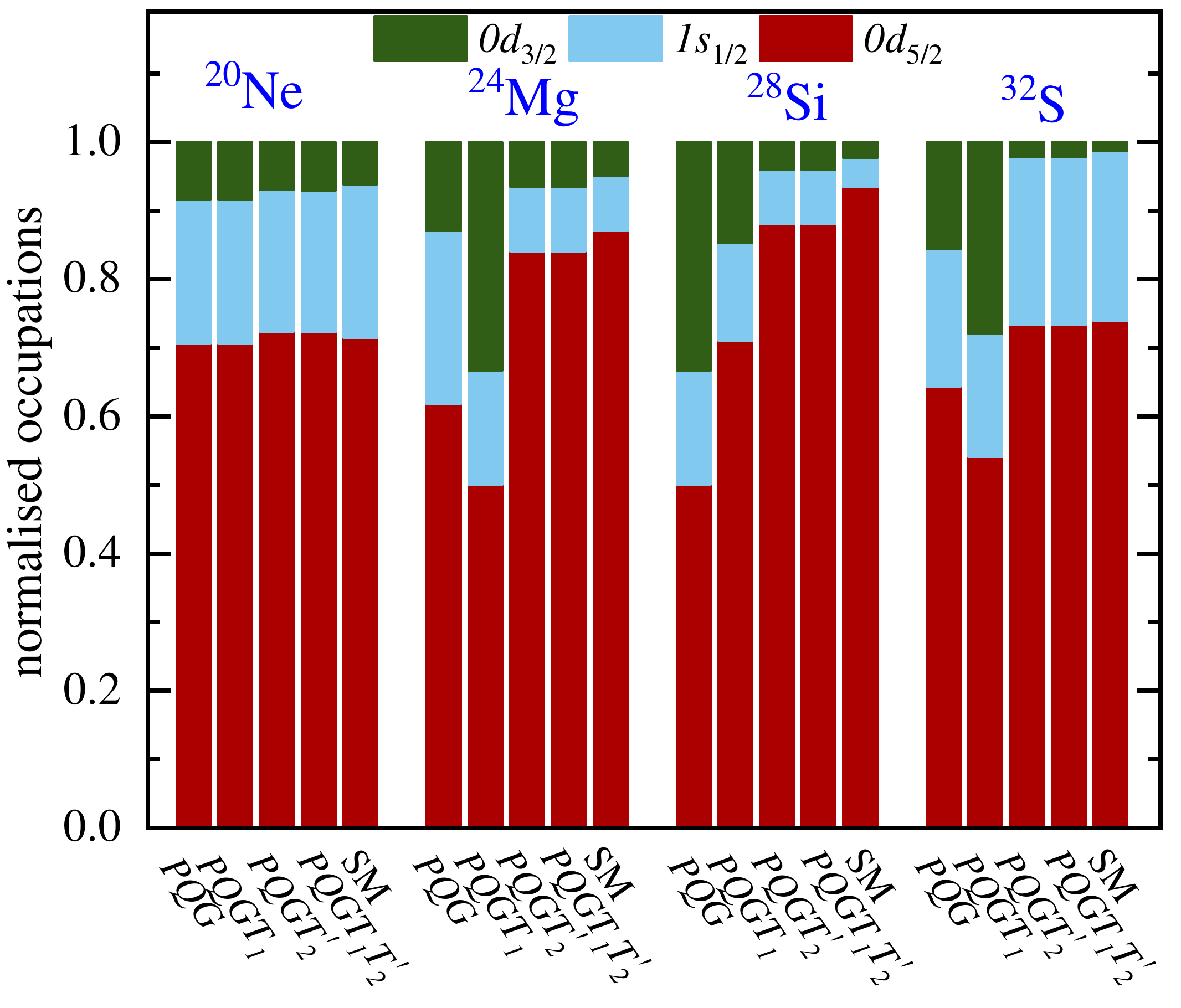}
\caption{Similar to Fig. \ref{USDB_occupation}, but where the USDB(1/2) interaction is utilized as Hamiltonian.}{\label{USDB_half_occupation}}
\end{figure}

Consequently, we also investigate the dependence on interaction strength by considering a softer nucleon-nucleon interaction. For this, we multiply the matrix elements of
the USDB effective interaction by 1/2, while keeping the same single-particle energies.
We denote this interaction by USDB(1/2). As the contribution of the USDB(1/2) residual two-body interaction is much smaller than that of the initial USDB interaction, we are able to quantify the error induced by the approximate \textit{N-representability} character of the 2RDM issued from energy minimization. Clearly, the USDB(1/2) interaction can only be used for benchmarking purposes and is not suitable to describe $sd$-shell nuclei.
We then reconsider the same nuclei with the variational 2RDM calculations with the USDB(1/2) interaction as those with the initial USDB interaction, i.e.,  the $N=Z$ $sd$-shell nuclei $^{20}$Ne, $^{24}$Mg, $^{28}$Si, and $^{32}$S.

The calculated energies of these nuclei using the variational 2RDM and shell-model method are shown in Fig. \ref{USDB_half_E} and Tab. \ref{USDB_half_Table}.
Even when using a softer nucleon-nucleon interaction, the variational 2RDM calculations with $PQG$ condition still cannot provide an adequate description of those nuclei. Indeed, the obtained energies are overbound when compared with the exact values by up to about 45 MeV.
Similar to the previous studies, the $T_1$ condition gives only but a small contribution to binding energies with the USDB(1/2) interaction, while the inclusion of the $T_2^\prime$ condition largely improves the calculations.
Moreover, the results  obtained with $PQG T_2^\prime$ and $PQG T_1 T_2^\prime$  are nearly the same based on the USDB(1/2) interaction.
The largest energy discrepancy found in the binding energies of the considered nuclei using the USDB(1/2) 2RDM with $PQG T_1 T_2^\prime$ conditions is about 1.8 MeV. This value is smaller than that obtained with the USDB interaction in the same conditions, where it is about 6.5 MeV.
The calculated occupations in the variational 2RDM using USDB(1/2) interaction follow a similar trend as that with the USDB interaction. (see Fig. \ref{USDB_half_occupation}). Due to the soft character of the USDB(1/2) interaction, the occupations obtained in the variational 2RDM calculations with $PQG T_1 T_2^\prime$ or $PQG T_2^\prime$ conditions are nearly the same as those provided by SM calculations.

\section{Towards a practical variational 2RDM method for atomic nuclei}
The error induced by the approximate \textit{N-representability} of the variational 2RDM strongly depends on the correlations induced by the two-body interaction of the considered Hamiltonian. As mentioned in Ref.\cite{Verstichel_PhD}, the 1RDM provided by the variational RDM calculation is always exactly \textit{N-representable}, because this property is equivalent to have the 1RDM eigenvalues to be between 0 and 1. Consequently, the ground-state energy arising from a one-body Hamiltonian has to be exactly reproduced in a variational RDM calculation. Conversely, overbinding arises when including two-body forces, which increases along with the strength of the interaction. While this discrepancy is mild in atomic and molecular calculations in the presence of $PQG$ conditions, errors can still reach several MeVs in nuclear binding energies by using $PQG T_1 T_2'$ constraints, even after making the nuclear interaction softer by a factor of two. Indeed, the maximal error is reduced  by about a factor of 3.5 after dividing the matrix elements of the USDB interaction by 2, i.e.,~they are of the same order of magnitude.

Therefore, a direct implementation of the variational 2RDM method cannot be used in practice for atomic nuclei, contrary to atomic and molecular systems, where it can directly provide almost exact binding energies. While the situation has been much improved since the original work of Mihailovi{\'c} and Rosina of Ref.\cite{MIHAILOVIC1975221}, it is far from being satisfactory, as the typical error made on nuclear binding energies in systematic calculations such as density functional theory (DFT) \cite{RevModPhys.75.121,PhysRevC.82.024313,PhysRevC.97.044313} and microscopic-macroscopic models \cite{PhysRevC.86.044316,PhysRevLett.108.052501} is typically 0.5 to 1 MeV. Clearly, one has to devise alternative strategies in order  to develop a variational 2RDM method which can be of practical interest in nuclear physics.

As mentioned in Sec.(\ref{Results}), one of the most interesting paths for that matter would be to use the variational principle in small model spaces \cite{PhysRevLett.105.213003}. For example, one can diagonalize exactly the Hamiltonian matrix if one takes into account only a few basis orbitals. The obtained energy provides with a useful constraint for the exact ground state. This method has been implemented, and one has seen that it accelerates convergence significantly. However, it has been noticed during the first tests that the constraint works well only if the correlation energy provided by the rest of the model space is of a few hundreds of keV at most. As consequence, constraints based on the variational principle are interesting mainly when eigenstates are built with a very large part from a few configurations of low energy. This is especially the case in the Gamow shell model \cite{PhysRevLett.89.042501,PhysRevLett.89.042502,0954-3899-36-1-013101,PhysRevC.102.034302}, where continuum coupling is small, so that the $0p$ or $1s0d$ orbitals are mainly occupied \cite{0954-3899-36-1-013101}. Moreover, according to the results obtained in Sec.(\ref{COSM_RDM}) for $^{6,8}$He and $^8$Be, it might also be possible that $PQG$ constraints are sufficient to impose \textit{N-representability} in the presence of constraints using the variational principle in small model spaces.

The situation is reversed when one enters the zones of the nuclear chart associated to medium and heavy nuclei. From a numerical point of view, the variational 2RDM method is well suited for ground states therein, because one typically has few basis orbitals but many valence nucleons in model space. Consequently, it is possible to apply in practice the $PQG T_1 T_2^\prime$ constraints. Moreover, the variational 2RDM method can be applied easily  to nuclei having an even or odd number of protons or neutrons, contrary to the Hartree-Fock-Bogoliubov method, for example, which becomes cumbersome to apply with odd nuclei \cite{RevModPhys.75.121}. Contrary to light nuclei, however, the main problem here is that the nuclear interaction acts similarly on all basis orbitals, and shell-model wave functions spread in a large configuration space \cite{RevModPhys.77.427}. Hence, the variational 2RDM method may lead to overbound ground states. In fact, the calculations depicted in Sec.(\ref{USDB_RDM}) have shown that this is already the case in the $sd$ shell.  Consequently, it might be of interest here to minimize the ground-state energies of Hamiltonian whose two-body part is reduced. The overbinding caused by the approximate \textit{N-representability} of the 2RDM issued from energy minimization would then be compensated by the use of smaller two-body matrix elements. The 2RDM obtained by energy minimization could then become closer to that of the physical ground state, even though it would arise from the use of a different Hamiltonian. This procedure can be related to the Hohenberg-Kohn and Kohn-Sham theories \cite{PhysRev.136.B864,PhysRev.140.A1133}, at the origin of DFT \cite{RevModPhys.75.121}, where the nuclear density, as a function of the correlated ground state, is determined from a mean-field functional. The validity of this method could be checked in shell-model spaces where diagonalization of Hamiltonian can be effected, or by using solvable Hamiltonians for that matter, of SU(3) type for example \cite{Elliott_SU3_1,Elliott_SU3_2}.

\section{Summary} \label{Summary}

We have developed the variational two-particle reduced density matrix method for atomic nuclei, and have applied it to the ground states of nuclear systems. The $N \text{-}representability$ conditions, represented by the two-particle $P$, $Q$, and $G$ metric matrices, as well as the partial three-particle $T_1$ and $T_2^{\prime}$ metric matrices, have been derived and applied for the first time  within the $J$-scheme formalism.  In addition, the neutron and proton numbers $N$ and $Z$, total spin $J$ and the isospin $T$ are considered as constraints in our calculations.

We have firstly applied the variational two-particle reduced density matrix method within the cluster-orbital shell-model framework. This has allowed to use a phenomenological interaction, which is convenient for benchmarking purposes. The $^{6,8}$He and $^8$Be ground-state energies have been calculated using $PQG$ constraints, and compared with shell-model calculations, which can be considered as the exact result (since shell-model dimensions are tractable). The results show that the two-particle reduced density matrix calculations could provide good descriptions of ground-state energies and partial-wave occupations in $^{6,8}$He. However, in the case of $^8$Be, the partial-wave occupations exhibits a clear difference from the shell-model calculation and a small but noticeable overbinding of ground-state energy.  Due to the large number of basis orbitals in the cluster-orbital shell-model  model space, only the $PQG$ conditions could be imposed, because computational cost becomes too large if one includes the partial 3-positive $T_1$ and $T_2^{\prime}$ conditions. Consequently, one had to consider a different Hamiltonian and shell-model space to study the impact of the $T_1$ and $T_2^{\prime}$ constraints.

We have then employed the variational two-particle reduced density matrix method using the standard USDB interaction in the $sd$ shell. Indeed, the number of basis orbitals is small, on the one hand, and Hamiltonian matrices can always be diagonalized exactly using the shell-model framework. Hence, we could perform variational two-particle reduced density matrix calculations with all possible sets of conditions, i.e.,~$PQG$, $PQG T_1$,  $PQG T_2^\prime$, and $PQG T_1 T_2^\prime$. We firstly studied the oxygen even-mass isotopes. The results showed that the calculations with $PQG$ conditions lead to overbound ground-state energies. The situation is largely improved when the  $T_2^\prime$ condition is imposed.  The variational two-particle reduced density matrix calculations with $PQG T_2^\prime$ and $PQG T_1 T_2^\prime$ can then provide an accurate description of oxygen isotopes. The largest difference of ground-state energies compared to exact values is about 100 keV, so that calculations using the variational two-particle reduced density matrix method can be considered as optimal. Furthermore, the obtained occupations with $PQG T_2^\prime$ or $PQG T_1 T_2^\prime$ conditions are nearly the same as those arising from shell-model calculations.

However, the calculations of $N=Z$ even-even mass isotopes in that same framework exhibit a large discrepancy with exact results, even in the presence of $PQG T_2^\prime$ or $PQG T_1 T_2^\prime$ constraints. The largest error is indeed as large as 6.5 MeV, which occurs in $^{28}$Si.
In order to investigate the effect of Hamiltonian correlations in ground-state energy errors,
we made the USDB Hamiltonian softer by halving its two-body matrix elements.
The error in the ground-state energy of $^{28}$Si is then decreased to 1.8 MeV with $PQG T_2^\prime$ or $PQG T_1 T_2^\prime$ conditions, which is smaller than with the initial USDB interaction, but remains too large nevertheless compared to those obtained with other methods to solve the nuclear many-body problem.

These results then suggest that the present variational two-particle reduced density matrix method cannot be applied satisfactorily to  nuclear systems as in atomic or molecular physics. Indeed, it is impossible to reach the same precision for binding energy due to the strength of inter-nucleon correlations, even if one uses $PQG T_1 T_2^\prime$ constraints. Theoretical frameworks have then been proposed, which will be the object of future works. The use of the variational principle in reduced model spaces would then provide with an additional stringent condition in light nuclei, because the most important configurations of their ground states are built with the orbitals of lowest energy.
Therefore, it would be more interesting to optimize the 2RDM of a less correlated Hamiltonian, so that the energy overbinding arising from the approximate \textit{N-representability} of the 2RDM would be mitigated by the use of a softer interaction.


\textit{Acknowledgments} --
This work has been supported by the National Key R\&D Program of China under Grant No. 2018YFA0404401; the National Natural Science Foundation of China under Grants No. 11835001,  11921006, 12035001, and  11975282; the State Key Laboratory of Nuclear Physics and Technology, Peking University under Grant No. NPT2020KFY13; the Strategic Priority Research Program of Chinese Academy of Sciences under Grant No. XDB34000000; and the CUSTIPEN (China-U.S. Theory Institute for Physics with Exotic Nuclei) funded by the U.S. Department of Energy, Office of Science under Grant No. de-sc0009971. We acknowledge the High-Performance Computing Platform of Peking University for providing computational resources.

\bibliography{Ref}

\end{document}